\documentclass[a4paper,11pt]{article}
\pdfoutput=1 

\usepackage{jheppub} 
\usepackage{graphics}
\usepackage{times}
\usepackage{color}
\usepackage{amsfonts,amssymb,stmaryrd,latexsym,amsmath}
\usepackage{bm}
\usepackage{enumerate}
\usepackage{enumitem}
\usepackage{array} 

\newcommand{\be}{\begin{equation}}
\newcommand{\ee}{\end{equation}}
\newcommand{\ba}{\begin{eqnarray}}
\newcommand{\ea}{\end{eqnarray}}
\newcommand{\non}{\nonumber}
\newcommand{\al}{&}

\newcommand{\Tr}{\textrm{Tr}}

\newcommand{\order}[1]{\mathcal{O}\left(\delta^{#1}\right)}

\title{Baryon electric dipole moments from strong CP violation}

\author[a]{Feng-Kun Guo}
\author[a,b]{and Ulf-G.~Mei{\ss}ner}
\affiliation[a]{Helmholtz-Institut f\"ur Strahlen- und Kernphysik and Bethe
   Center for Theoretical Physics,
   Universit\"at Bonn,  D-53115 Bonn, Germany }
\affiliation[b]{Institute for Advanced Simulation, Institut f\"{u}r Kernphysik
   and J\"ulich Center for Hadron Physics,
   JARA-FAME and JARA-HPC, Forschungszentrum J\"{u}lich,
   D-52425 J\"{u}lich, Germany}
\emailAdd{fkguo@hiskp.uni-bonn.de}
\emailAdd{meissner@hiskp.uni-bonn.de}

\abstract{
 The electric dipole form factors and moments of the ground state baryons
are calculated in chiral perturbation theory at next-to-leading order. We show that
the baryon electric dipole form factors at this order depend only on two
combinations of low-energy constants. We also derive various relations that are
free of unknown low-energy constants. We use recent lattice QCD data to calculate
all baryon EDMs. In particular, we find $d_n = -2.9\pm 0.9$ and $d_p = 1.1\pm 1.1$
in units of $10^{-16}\,e\,\theta_0$~cm. Finite volume corrections to the electric
dipole moments are also worked out. We show that for a precision
extraction from lattice QCD data, the next-to-leading order terms have to be
accounted for.
}

\keywords{Chiral Lagrangians, CP violation, Lattice QCD}

\begin{document}
\maketitle
\flushbottom

\section{Introduction}

The neutron electric dipole moment (EDM) is a sensitive probe of CP
violation in the Standard Model (SM) and beyond. The current experimental
limit $|d_n| \leq 2.9 \cdot 10^{-26}\, e \,$cm~\cite{Baker:2006ts}
is still orders of magnitude larger than
the SM prediction due to weak interactions. Furthermore, in quantum
chromodynamics (QCD)
the breaking of the U(1)$_A$ anomaly allows for strong CP violation,
which is parameterized through the vacuum angle $\theta_0$. Therefore,
an upper bound on $d_n$ allows to constrain the magnitude of  $\theta_0$.
Furthermore, such electric dipole moments are very sensitive to physics
beyond the SM, see e.g.~\cite{Pospelov:2005pr}.
Many extensions of the SM in fact lead to larger EDMs than the tiny
SM predictions, so that any limit of
$d_n$ leads to bounds on the scale of the new physics. In this paper,
we concentrate on the CP violation generated by the $\theta$-term
of QCD.  New and on-going experiments with ultracold neutrons
strive to improve the aforementioned bounds even further, see
e.g.~\cite{LaGo} for a recent review. Furthermore, there are new experimental
proposals to measure the EDM of the proton and the deuteron in storage rings,
which in principle allow for an even higher sensitivity than obtained
with the instable neutron,
see e.g. refs.~\cite{Farley:2003wt,Semertzidis:2011qv,Rathmann:2011zz,Lehrach:2012eg}.

Besides these challenging experimental activities, first full lattice
QCD calculations of the neutron and the proton  electric dipole moment
are becoming available.  There exist three different methods of calculating
the nucleon EDM on the lattice. The EDM can be related
to the energy difference of the nucleon with different spin alignments in the presence
of an external electric field~\cite{Aoki:1989rx,Aoki:1990ix,Shintani:2006xr,Shintani:2008nt}.
It can also be obtained by calculating the electric dipole form factor (EDFF) at
finite momentum transfer $q^2$, see the definition in eq.~\eqref{eq:formfactors},
and extrapolating to the point with $q^2=0$~\cite{Shintani:2005xg,Berruto:2005hg},
\begin{equation}
   d_N = \frac{F_{3,N}(0)}{2m_N}.
\end{equation}
In addition, the nucleon EDM can also be calculated by analytically continuing
$\theta_0$ to a purely imaginary quantity~\cite{Izubuchi:2008mu,Aoki:2008gv}, as
the QCD action in the presence of the $\theta$-term becomes real in Euclidean space. For
a brief review of these methods, see ref.~\cite{confX}.
We do not discuss here another method~\cite{Gerrit} that relates
certain Fock state components of the EDM to the
Fock state expansion of the magnetic moment (derived in
light-front QCD)~\cite{Brodsky:2006ez} since it
is not clear what these relations imply for the
observable quantities~\cite{Liu:2008gr,Mereghetti:2010tp}.

These lattice studies require a careful study of the quark mass dependence
of the nucleon EDM to connect to the physical light quark masses. In addition, CP-violating
atomic effects can be sensitive to the nuclear Schiff moment, which
receives a contribution from the radius of the nucleon electric dipole
form factor (EDFF), see e.g.~\cite{Thomas:1994wi}.
It is thus of paramount interest to improve the existing
calculations of these fundamental quantities in the framework of
chiral perturbation theory (CHPT). In~\cite{Borasoy:2000pq}, the electric
dipole moments of the neutron and the $\Lambda$ were calculated within the
framework of U(3)$_L\times$U(3)$_R$ heavy-baryon chiral perturbation theory
and an estimate for  $\theta_0$ was given (for earlier works utilizing
chiral Lagrangians, see~\cite{Crewther:1979pi,Pich:1991fq,Narison:2008jp}).
In~\cite{Hockings:2005cn}, the
electric dipole form factor of the nucleon was analyzed to leading one-loop
accuracy in chiral SU(2). In that calculation, the form factor
originates entirely from the pion cloud. The strength of the form factor
was shown to be proportional to a non-derivative and CP-violating
pion--nucleon coupling $\bar g_{\pi NN}$ that was estimated from
dimensional analysis in~\cite{Hockings:2005cn}. In ref.~\cite{Ottnad:2009jw}, the results
of refs.~\cite{Borasoy:2000pq,Hockings:2005cn} were extended
to higher order based on a covariant version of U(3)$_L\times$U(3)$_R$
 baryon CHPT, and an expression for
$\bar g_{\pi NN}$ was given in terms of the measurable quantities. For other recent work on these
issues, see \cite{Mereghetti:2010tp,Mereghetti:2010kp}. Here, we extend
the studies of ref.~\cite{Ottnad:2009jw} to the baryon octet and make
contact to recent lattice QCD studies.

Furthermore, the leading contributions to the neutron EDM at finite volume and in
partially-quenched calculations were considered
in~\cite{O'Connell:2005un}, and in~\cite{Chen:2007ug} the leading
order extrapolation formula using a mixed action chiral Lagrangian is given.
Here, we work out the finite volume expression for the whole baryon octet
at next-to-leading order (NLO). In particular, we show that the formally suppressed
NLO corrections are sizeable, and even dominate for some
baryons due to large cancellations at leading order.

The manuscript is organized as follows. The underlying effective chiral
U(3)$_L\times$U(3)$_R$ Lagrangian is given in section~\ref{sec:lagr}. In
section~\ref{sec:BEDMs} we work out the baryon EDFFs and EDMs in the infinite volume
for varying quark masses. We derive new relations for all baryon EDMs
and show that these only depend on two combinations of unknown low-energy
constants (LECs). Using recent lattice data, we can give predictions for all
baryon EDMs. Then, in section~\ref{sec:fv}, we calculate the finite volume
corrections for the baryons at NLO and show that these
NLO corrections are substantial and must be included in any extraction from
lattice data. We end with a summary and outlook in section~\ref{sec:sum}.
Various technical aspects of our calculations are displayed in the appendices.

\section{Effective Lagrangian for strong CP violation}
\label{sec:lagr}

The most general gauge-invariant and renormalizable Lagrangian for QCD is
\begin{equation}
  \mathcal{L}_{QCD} = -\frac14 G_{\mu\nu}^{a} G^{a, \mu\nu} +
  \bar q \left( i \not\!\!{D} - \mathcal{M} \right) q + \theta
  \frac{g^2}{32\pi^2} G_{\mu\nu}^{a} \tilde G^{a, \mu\nu}  \,\,\, (a = 1,
  \ldots , 8) ~ ,
  \label{eq:QCDLagrangian}
\end{equation}
with $G_{\mu\nu}^a$ the gluon field strength tensor and
$\tilde G_{\mu\nu}^a = \varepsilon_{\mu\nu\lambda\sigma} G^{a, \lambda\sigma}/2$
its dual, $q$ collects the quark fields of various flavors, $D_\mu$ is the
gauge-covariant derivative, and $\mathcal{M}$ is the 
quark mass matrix.
The last term is the so-called $\theta$-term, which breaks the P and CP symmetries. It is
a consequence of the U(1)$_A$ anomaly. Because the theta-term is related to
chiral U(1) transformations of the quark fields (see, for instance, ref.~\cite{Donoghue}),
only the combination
\begin{equation}
   \theta_0 = \theta + \arg\det \mathcal{M}
\end{equation}
is a measurable quantity. In CHPT, one may
treat the $\theta$-term using an external field $\theta(x)$, and the QCD Green
functions can be obtained by expanding the generating functional around
$\theta(x)=\theta_0$ with real quark masses. Under an axial U(1)
transformation, one has
\begin{equation}
   \theta(x) \to \theta(x) - 2 N_f \alpha,
\end{equation}
where $N_f$ is the number of flavors, and $\alpha = (\alpha_R-\alpha_L)/2$. In the
limit of infinitely large number of colors $N_c$, the U(1)$_A$ anomaly is
absent. In this case, the spontaneous chiral symmetry breaking of
U(3)$_L\times$U(3)$_R$, which is a symmetry
of the QCD Lagrangian, into U(3)$_V$ gives nine Goldstone
bosons (the SU(3) flavor octet $\{\pi^\pm, \pi^0, K^\pm, K^0,\bar K^0, \eta_8\}$ and the
flavor singlet $\eta_0$).
Collecting these fields in $\tilde U(x)$, this transforms under
the axial U(1) transformation as
\begin{equation}
   \tilde U(x) \to  e^{i\alpha_R} \tilde U(x) e^{-i\alpha_L}.
\end{equation}
Thus, the combination
\begin{equation}
  \bar\theta(x) = \theta(x) - i \ln \det \tilde U(x) ,
\end{equation}
is invariant under chiral transformations.

With $\bar\theta(x)$, the most general chiral effective Lagrangian which is
invariant under U(3)$_L\times$U(3)$_R$ can be constructed. The original construction
at  order $\order{2}$ can be found in~\cite{Gasser:1984gg}, where
$\mathcal{O}(\delta)=\mathcal{O}(M_\phi,k)$ with $M_\phi$ and $k$ denoting the
Goldstone boson masses and a small momentum, respectively. Further, we count
$1/N_c$ as $\order{2}$.
Here, we adopt the notation used in~\cite{Borasoy:2000pq}, whose
formulation is partially based on
refs.~\cite{Leutwyler:1996sa,HerreraSiklody:1996pm}.
The most general chiral effective Lagrangian for mesons to second chiral order,
complying with the U(3)$_L \times$U(3)$_R$ symmetry, reads
\begin{eqnarray}
 \mathcal{L} \al=\al - V_0 + V_1 \,\Tr \bigl[ \nabla_\mu \tilde U^\dag \nabla^\mu
 \tilde U \bigr] + V_2 \,\Tr \bigl[ \tilde\chi^\dag \tilde U + \tilde\chi \tilde U^\dag \bigr]
+ i V_3 \,\Tr \bigl[ \tilde\chi^\dag \tilde U - \tilde\chi \tilde U^\dag \bigr] \nonumber \\
\al\al + V_4 \,\Tr \bigl[ \tilde U \nabla_\mu \tilde U^\dag \bigr] \Tr \bigl[ \tilde
U^\dag \nabla^\mu \tilde U \bigr] + V_5 \,\Tr \bigl[ \nabla_\mu \theta \nabla^\mu \theta  \bigr] ,
\label{eq:MesonicLagrangianVEVNotFixed}
\end{eqnarray}
where  $\tilde\chi = 2 B_0 M_q$ with $M_q=\text{diag}(m_u,m_d,m_s)$ the real quark
mass matrix, $\nabla_\mu \tilde U = \partial_\mu \tilde U - i r_\mu \tilde U + i
\tilde U l_\mu$, and the $V_i$'s are functions of $\bar\theta(x)$.

In order to use the above Lagrangian, we need to determine the vacuum. Denoting the
vacuum expectation value of $\tilde U$ by $U_0$, we can decompose $\tilde U$ as
\begin{equation}
   \tilde U = \sqrt{U_0} U \sqrt{U_0},
\end{equation}
where
\begin{equation}
 U = \exp \bigg( i \sqrt{\frac{2}{3}} \frac{\eta_0}{F_0}  +  i\frac{\sqrt{2}
 }{F_\pi} \phi\bigg) .
\end{equation}
with
\begin{equation}
   \phi = \left(
   \begin{array}{ccc}
      \frac{1}{\sqrt{2}}\pi^0 + \frac{1}{\sqrt{6}}\eta_8 & \pi^+ & K^+\\
      \pi^- & -\frac{1}{\sqrt{2}}\pi^0 + \frac{1}{\sqrt{6}}\eta_8 & K^0\\
      K^- & \bar{K}^0 & - \frac{2}{\sqrt{6}}\eta_8
   \end{array}
   \right).
\end{equation}
When $\theta_0=0$, the vacuum is trivial, $U_0=1$. When the $\theta_0$ angle is
switched on, the vacuum is shifted, and $U_0$ has to be determined by minimizing the
zero modes of the Lagrangian. One may parametrize the vacuum as
\begin{equation}
   U_0 = \text{diag} \left( e^{-i \varphi_u}, e^{-i \varphi_d}, e^{-i
   \varphi_s}\right).
\end{equation}
After the vacuum alignment,
one obtains the effective Lagrangian~\cite{Borasoy:2000pq}~\footnote{In
ref.~\cite{Borasoy:2000pq}, there is one more term $ i (V_3 - \mathcal{B} V_{2} ) \Tr[\chi
\left(U-U^\dag\right)]$. However, this term vanishes exactly because $\mathcal{B}$ is
defined as $V_3/V_2$.
}
\begin{eqnarray}
\mathcal{L}_{\phi} &=& - V_0 + V_1\, \Tr \left[ \nabla_\mu U^\dag
\nabla^\mu U \right] + \left( V_{2} + \mathcal{B} V_3 \right) \Tr
\left[ \chi \left( U + U^\dag \right) \right] - i\mathcal{A} V_{2}\, \Tr
\left[ U - U^\dag\right] \nonumber \\
&& + \mathcal{A} V_3\, \Tr \left[ U + U^\dag
\right] + V_4\, \Tr \left[ U \nabla_\mu U^\dag \right] \Tr \left[
U^\dag \nabla^\mu U \right] ,
\label{eq:MesonLagrangian}
\end{eqnarray}
where $\chi=2B_0 \text{diag}(m_u\cos\varphi_u,m_d\cos\varphi_d,m_s\cos\varphi_s)$,
and $\mathcal{A}$, $\mathcal{B}$ are complicated functions
of the $V_i$'s, see e.g.~ref.~\cite{Borasoy:2000pq}. Introducing the notation
$\bar\theta_0=\theta_0-\sum_q \varphi_q$, one may expand the $V_i$'s in terms of
$\bar\theta_0$.
All of them except for $V_3$, which is an odd function, are even functions of
$\bar\theta_0$.
In the leading approximation, $\mathcal{A}$ and $\mathcal{B}$ are given by
\begin{equation}
\mathcal{A} = \frac{V_0^{(2)}}{V_2^{(0)}} \bar\theta_0 + \order{4} , \qquad
\mathcal{B} = \frac{V_3^{(1)}}{V_2^{(0)}} \bar\theta_0 + \order{6}.
\end{equation}
Since $$\ln\det U= \Tr \ln U = i \frac{\sqrt{6}}{F_0} \eta_0,$$
the $V_{i}$ are now
functions of $\bar\theta_0 + \sqrt{6}\eta_0 / F_0 $. The correct
normalization of the kinetic terms of the Goldstone boson fields can be obtained by
requiring
\begin{equation}
   V_1(0) = V_2(0) = \frac{F_\pi^2}{4}, \qquad V_4(0) = \frac1{12}\left( F_0^2 -
   F_\pi^2 \right).
\end{equation}
Finally, the quantity $\bar \theta$ can be
expressed in terms of the measurable quantity $\theta_0$~\cite{Ottnad:2009jw},
\begin{equation}
 \bar\theta_0 = \left[ 1 + \frac{4 V_0^{(2)}}{F_\pi^2}
\frac{4 M_K^2 - M_\pi^2}{M_\pi^2\left(2 M_K^2 - M_\pi^2 \right)}\right]^{-1}
\,\theta_0 . 
\label{eq:theta0barMK}
\end{equation}
One sees that $\bar\theta_0 = \order{2}$, because $V_0^{(2)}$ is of the zeroth chiral
order.

Similarly, one can also construct the most general effective Lagrangian in
U(3)$_L \times$U(3)$_R$ CHPT for the baryon octet (for a general discussion of
effective Lagrangians for the $\theta$-term, see \cite{Mereghetti:2010tp})
\begin{equation}
   B = \left(
   \begin{array}{ccc}
      \frac1{\sqrt{2}}\Sigma^0 + \frac1{\sqrt{6}}\Lambda & \Sigma^+ & p\\
      \Sigma^- & -\frac1{\sqrt{2}}\Sigma^0 + \frac1{\sqrt{6}}\Lambda & n\\
      \Xi^- & \Xi^0 & - \frac{2}{\sqrt{6}}\Lambda
   \end{array}
   \right).
\end{equation}
The Lagrangian up to the second chiral order is (only the terms relevant
to our calculation are displayed; for details, see ref.~\cite{Borasoy:2000pq})
\begin{eqnarray}
  \mathcal{L}_{\phi B} \al=\al i \,\Tr \bigl[ \bar{B} \gamma^\mu [ D_\mu , B ] \bigr]
  - \mathring{m} \,\Tr [ \bar{B} B ]
  - \frac{D}{2} \Tr \bigl[ \bar{B} \gamma^\mu \gamma_5 \{ u_\mu, B \} \bigr] -
  \frac{F}{2} \Tr \bigl[ \bar{B} \gamma^\mu \gamma_5 [ u_\mu, B ] \bigr] \nonumber \\
  \al\al + \frac{w_0}{2} \,\Tr \bigl[ \bar{B} \gamma^\mu \gamma_5 B \bigr] \Tr [
  u_\mu ] + b_D \,\Tr \bigl[ \bar{B} \left\{ \tilde \chi_+, B
 \right\} \bigr] + b_F \,\Tr \bigl[ \bar{B} \left[ \tilde\chi_+, B
 \right] \bigr] + b_0 \,\Tr [ \bar{B} B ] \,\Tr \left[ \tilde\chi_+\right] \nonumber \\
 \al\al
 +4 \mathcal{A} \,w'_{10} \frac{\sqrt{6}}{F_0} \eta_0 \Tr [ \bar{B} B ] +
 i \Big( w'_{13} \bar\theta_0 + w_{13} \frac{\sqrt{6}}{F_0} \eta_0 \Big)
\Tr \bigl[ \bar{B} \sigma^{\mu\nu} \gamma_5 \left\{ F_{\mu\nu}^{+} , B \right\}
\bigr]\nonumber\\
 \al\al
+ i \Big( w'_{14} \,\bar\theta_0 + w_{14} \frac{\sqrt{6}}{F_0} \eta_0 \Big)
\Tr \left[ \bar{B} \sigma^{\mu\nu} \gamma_5 \left[ F_{\mu\nu}^{+} , B \right]
\right],
 \label{eq:BaryonLagrangian}
\end{eqnarray}
where $\tilde\chi_+ = \chi_+ - i \mathcal{A} (U-U^\dag)$,
and we use the same notation $w_{10}'=w_{10}+3 w_{12}/2$ as in ref.~\cite{Ottnad:2009jw}
with the low-energy constants (LECs) $w_{10}$ and $w_{12}$
defined in ref.~\cite{Borasoy:2000pq}. We remark here that although there seems
to be quite a number of unknown LECs, i.e. $w_0, w_{10}', w_{13}, w'_{13},
w_{14}$ and $w'_{14}$, only {\it two} combinations of these will
finally appear in
the expressions of the baryon EDFFs.

\section{Baryon electric dipole form factors in the infinite volume}
\label{sec:BEDMs}

The electromagnetic form factors of a baryon are defined by
\begin{eqnarray}
  \langle B(p') | J_{\rm em}^\nu | B(p) \rangle \al=\al \bar u(p') \left[
  \gamma^{\nu}F_1\left(q^2\right) - \frac{i\, F_2\left(q^2\right)}{2m_B}
  \sigma^{\mu\nu}q_\mu \right. \nonumber\\
   \al\al \left. + i \left( \gamma^{\nu}q^2\gamma_5 - 2m_B
   q^{\nu}\gamma_5\right)F_A\left(q^2\right) - \frac{F_3\left(q^2\right)}{2m_B}
   \sigma^{\mu\nu}q_\mu\gamma_5 \right] u(p)~,
\label{eq:formfactors}
\end{eqnarray}
where $F_1(q^2)$ and $F_2(q^2)$ are the P- and CP-conserving Dirac and Pauli form
factors, respectively, while $F_A(q^2)$ is the P-violating anapole form factor, and
$F_3(q^2)$ if the electric dipole form factor (EDFF) which breaks P and CP
symmetries. Here, $J_{\rm em}^\nu$ is the electromagnetic current, the baryon mass
is denoted by $m_B$, and $q_\mu = p'_\mu-p_\mu$ is the four-momentum transfer.

In what follows, we will consider the dipole form factor $F_3(q^2)$.
The electric dipole moment of a baryon is defined as the electric
dipole form factor at $q^2=0$
\begin{equation}
d_{B} = \frac{F_{3,B}(0) }{2m_B}.
\label{eq:EDM_def}
\end{equation}
Note that similar to the case of the neutron electric form factor, we do
not include the normalization of the form factor at $q^2 = 0$ in this
definition.

\subsection{Baryon electric dipole form factors up to NLO}
\label{sec:EDFF}

\begin{figure}[t]
\centering
\includegraphics[width=\linewidth]{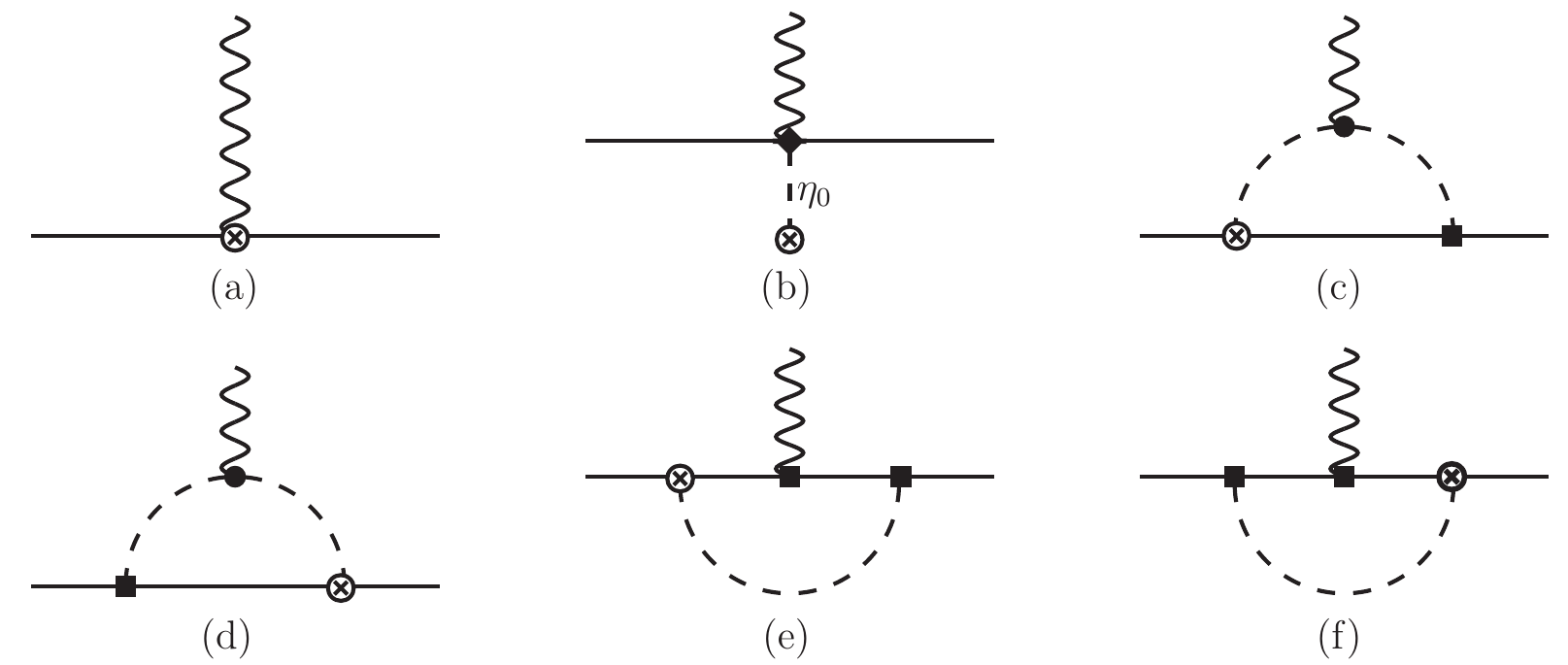}
\caption{Feynman diagrams contributing to the baryon EDFFs up to next-to-leading
order, where ${\bm \otimes}$ denotes a CP violating vertex, black dots represent
the second order mesonic vertices, filled squares and diamonds are the first and
second order baryonic vertices, respectively. \label{fig:feyn}}
\end{figure}

\begin{table}[t]
\begin{center}
   \begin{tabular}{|l | c |} \hline
      Baryons & Tree-level \\ \hline
      $p$ & $-\frac{4}{3} e \bar{\theta }_0 \left[\alpha  \left(w_{13}+3
      w_{14}\right)+w_{13}'+3 w_{14}'\right]$ \\
      $n$ & $\frac{8}{3} e \bar{\theta }_0 \left(\alpha  w_{13}+w_{13}'\right)$ \\
      $\Lambda$  & $\frac{4}{3} e \bar{\theta }_0 \left(\alpha
         w_{13}+w_{13}'\right)$      \\
      $\Sigma ^+$ & $-\frac{4}{3} e \bar{\theta }_0 \left[\alpha  \left(w_{13}+3
         w_{14}\right)+w_{13}'+3 w_{14}'\right]$ \\
      $\Sigma ^0$ & $-\frac{4}{3} e \bar{\theta }_0 \left(\alpha
         w_{13}+w_{13}'\right)$  \\
      $\Sigma ^-$ & $-\frac{4}{3} e \bar{\theta }_0 \left[\alpha  \left(w_{13}-3
         w_{14}\right)+w_{13}'-3 w_{14}'\right]$ \\
      $\Xi ^0$ & $\frac{8}{3} e \bar{\theta }_0 \left(\alpha  w_{13}+w_{13}'\right)$  \\
      $\Xi ^-$ & $-\frac{4}{3} e \bar{\theta }_0 \left[\alpha  \left(w_{13}-3
         w_{14}\right)+w_{13}'-3 w_{14}'\right]$ \\
      \hline
   \end{tabular}
   \caption{\label{tab:tree} Tree-level contribution to the EDFFs of the octet
   baryons.}
\end{center}
\end{table}
Consider first the tree level contributions to the baryon EDFFs,
figure~\ref{fig:feyn}(a,b). In particular, these feature the counterterms
$w_{13,14}'$ and $w_{13,14}$  at $\order{2}$~\cite{Borasoy:2000pq,Ottnad:2009jw}.
The expressions of the form factor $F_{3}(q^2)/(2 m)$ of the ground state octet
baryons from the tree-level diagrams are collected in
table~\ref{tab:tree},~\footnote{The overall sign of the tree-level expression for
the neutron EDM given in ref.~\cite{Ottnad:2009jw} should be positive.} where
$\alpha = 144 V_0^{(2)} V_3^{(1)}/(F_0 F_\pi M_{\eta_0})^2$. One notices that the
tree-level contributions to various baryons depend only on two combinations of the
LECs: $\alpha w_{13}+w_{13}'$ and $\alpha w_{14}+w_{14}'$. The dependence on the
latter may be written in the general form $-4 Q_B e\bar\theta_0(\alpha
w_{14}+w_{14}')$, with $Q_B$ the baryon electric charge.

There are four loop graphs contributing to the baryon EDFFs up to NLO, see
figure~\ref{fig:feyn}(c-f).
Formally, there are more diagrams. However, due to cancellations they
only contribute starting at  next-to-next-to-leading order. For details, see
refs.~\cite{Ottnad:2009jw,Ottnadthesis}. The expression for the sum of loops of a
given pair of any meson-baryon intermediate state is
\begin{eqnarray}
   \frac{F_{3}^{\,M\tilde m}(q^2)}{2 m} \al=\al \frac{8 e\bar\theta_0
   V_0^{(2)}}{F_\pi^4} \left\{ C_{\rm cd} \left[ -
   J_{MM}(q^2)+\left(2 m 
   (m-\tilde m)+M^2 -
   \frac{q^2}2\right) J_{MM\tilde m}(q^2,m^2) \right] \right. \nonumber\\
   \al\al + \left( C_{\rm cd} + C_{\rm ef} \right) J_{M\tilde m}(m^2) \bigg\} +
   \order{4},
   \label{eq:F3loop}
\end{eqnarray}
where $m$ is the mass of the external baryon, $M$ and $\tilde m$ are the masses of
the meson and baryon in the loops, respectively. We have made use of the fact that
the baryon mass difference is  $\order{2}$, see appendix~\ref{app:mb}. Keeping only
the leading order, the loop expression does not depend on the baryon mass, and reads
\begin{equation}
   \frac{F_{3\rm LO}^{\,M\tilde m}(q^2)}{2 m} = -\frac{8 e\bar\theta_0
   V_0^{(2)}}{F_\pi^4} C_{\rm cd} J_{MM}(q^2) .
   \label{eq:F3loopLO}
\end{equation}
The expressions for the loop functions involved in eqs.~\eqref{eq:F3loop} and
\eqref{eq:F3loopLO}, making use of infrared
regularization~\cite{Becher:1999he}, are collected in appendix~\ref{app:loop}.

\begin{table}[t]
\begin{center}
   \begin{tabular}{|l | c| c c |} \hline
      Baryons & Loops       & $C_{\rm cd}$       & $C_{\rm ef}$ \\
      \hline
      $n$ & $\{\pi^-,p\}$ & $2 (D+F) \left(b_D+b_F\right)$ & $-2 (D+F)
            \left(b_D+b_F\right)$ \\
         & $\{K^+,\Sigma^-\}$ & $-2 (D-F) \left(b_D-b_F\right)$ & $2
         (D-F) \left(b_D-b_F\right)$ \\
      \hline
      $p$ & $\{\pi^0,p\}$ & 0 & $-(D+F)\left(b_D+b_F\right)$      \\
          & $\{\pi^+,n\}$ & $-2(D+F)\left(b_D+b_F\right)$ & 0 \\
          & $\{K^0,\Sigma^+\}$ & 0 & $-2 (D-F) \left(b_D-b_F\right)$ \\
          & $\{K^+,\Lambda \}$ & $-\frac{1}{3} (D+3 F) \left(b_D+3 b_F\right)$
          & 0    \\
          & $\{K^+,\Sigma^0\}$ & $-(D-F)\left(b_D-b_F\right)$ & 0\\
          & $\{\eta_8,p\}$ & 0 & $-\frac13(D-3 F)\left(b_D-3 b_F\right)$ \\
          & $\{\eta_0,p\}$ & 0 & $-\frac{2F_{\pi}^2}{3 F_0^2} \beta$ \\
      \hline
   \end{tabular}
   \caption{\label{tab:CN} Possible loops contributing to the EDFFs of the nucleons
         and the corresponding coefficients $C_{\rm cd}$ and $C_{\rm ef}$. The
         intermediate states of the loops are listed in the second column for each nucleon.}
\end{center}
\end{table}

Different baryons receive contributions from loops with different intermediate
states. For instance, the loops for the neutron can be $\{ \pi^-, p \}$ and
$\{ K^+, \Sigma^- \}$, while they are $\{ \pi^+, n \}$, $\{ \pi^0(\eta_8,\eta_0), p \}$,
$\{ K^+, \Sigma^0(\Lambda) \}$ and $\{ K^0, \Sigma^+ \}$ for the proton. A list of
the possible intermediate states and the corresponding coefficients $C_{\rm cd}$ and
$C_{\rm ef}$ for the nucleons are given in table~\ref{tab:CN}, where we have defined
\begin{equation}
\beta = (2 D-3 w_0 ) \left(2 b_D+3 b_0+6 w_{10}'\right)~,
\end{equation}
for brevity. The loops and coefficients for the $\Sigma,\Lambda$ and $\Xi$ hyperons
are collected in table~\ref{tab:CH}. The loops are divergent. Up to $\order{3}$, the
divergences can be absorbed into the renormalization of $w_{13}'$ and $w_{14}'$,
\begin{eqnarray}
   w_{13}' \al=\al w_{13}^{\prime\,r}(\mu) + \frac{24 V_0^{(2)}}{F_\pi^4} \left(D
   b_F+F b_D\right) L, \nonumber \\
   w_{14}' \al=\al w_{14}^{\prime\,r}(\mu) + \frac{8V_0^{(2)}}{3F_\pi^4} \left(5D
   b_D+9F b_F\right) L,
\end{eqnarray}
where the divergence is contained in
\begin{equation}
L = \frac{\mu^{d-4}}{(4\pi)^2} \left\{
\frac1{d-4} - \frac12 \left[ \ln (4\pi) + \varGamma'(1) + 1 \right] \right\}~,
\end{equation}
with $d$ the number of space-time dimension, and $w_{13,14}^{\prime\,r}(\mu)$ are
the finite parts of $w_{13,14}'$. Notice that $w_{13,14}^{\prime\,r}(\mu)$ depend
on the renormalization scale $\mu$ through $L$. The scale dependence cancels with
that of the loops, and as a result, the final expressions of the EDFFs are
scale-independent. With the coefficients collected in tables~\ref{tab:CN} and
\ref{tab:CH}, one can easily obtain the explicit expressions of the baryon EDFFs up
to NLO. They are listed in appendix~\ref{app:expressions}.

\begin{table}[t]
\begin{center}
   \begin{tabular}{|l | c | c c |} \hline
      Baryons & Loops       & $C_{\rm cd}$       & $C_{\rm ef}$ \\
      \hline
      $\Sigma^+$ & $\{\pi^0,\Sigma^+\}$ & 0 & $-4 F b_F$ \\
         & $\{\pi^+,\Lambda \}$ & $-\frac{4}{3} D b_D$ & 0 \\
         & $\{\pi^+,\Sigma^0\}$ & $-4 F b_F$ & 0 \\
         & $\{\bar{K}^0,p\}$ & 0 & $-2 (D-F)\left(b_D-b_F\right)$ \\
         & $\{K^+,\Xi^0\}$ & $-2 (D+F) \left(b_D+b_F\right)$ & 0 \\
         & $\{\eta_8,\Sigma^+\}$ & 0 & $-\frac{4}{3} D b_D$ \\
         & $\{\eta_0,\Sigma^+\}$ & 0 & $-\frac{2 F_{\pi }^2}{3 F_0^2} \beta$ \\
      \hline
      $\Sigma^0$ & $\{\pi^+,\Sigma^-\}$ & $-4 F B_F$ & $4 F B_F$ \\
         & $\{\pi^-,\Sigma^+\}$ & $4 F B_F$ & $-4 F B_F$ \\
         & $\{K^+,\Xi^-\}$ & $- (D+F)\left(b_D+b_F\right)$ &
         $(D+F)\left(b_D+b_F\right)$ \\
         & $\{K^-,p\}$ & $(D-F)\left(b_D-b_F\right)$ & $-(D-F)\left(b_D-b_F\right)$ \\
      \hline
      $\Sigma^-$ & $\{\pi^0,\Sigma^-\}$ & 0 & $4 F b_F$ \\
         & $\{\pi^-,\Lambda \}$ & $\frac{4}{3} Db_D$ & 0 \\
         & $\{\pi^-,\Sigma^0\}$ & $4 F b_F$ & 0 \\
          & $\{K^0,\Xi^-\}$ & 0 & $2 (D+F) \left(b_D+b_F\right)$ \\
          & $\{K^-,n\}$ & $2 (D-F) \left(b_D-b_F\right)$ & 0 \\
          & $\{\eta_8,\Sigma^-\}$ & 0 & $\frac{4}{3} D b_D$ \\
          & $\{\eta_0,\Sigma^-\}$ & 0 & $\frac{2 F_{\pi }^2}{3 F_0^2} \beta$ \\
          \hline
      $\Lambda$ & $\{\pi ^+,\Sigma^-\}$ & $-\frac{4}{3}Db_D$ & $\frac{4}{3}Db_D$ \\
            & $\{\pi^-,\Sigma^+\}$ & $\frac{4}{3}Db_D$ & $-\frac{4}{3}Db_D$      \\
            & $\{K^+,\Xi^-\}$ & $-\frac13(D-3F) \left(b_D-3 b_F\right)$ &
            $\frac13(D-3 F) \left(b_D-3 b_F\right)$ \\
            & $\{K^-,p\}$ & $\frac13(D+3 F) \left(b_D+3 b_F\right)$ &
            $-\frac13(D+3 F) \left(b_D+3 b_F\right)$ \\
      \hline
       $\Xi^0$ & $\{\pi^+,\Xi^-\}$ & $-2(D-F)\left(b_D-b_F\right)$ &
            $2(D-F)\left(b_D-b_F\right)$ \\
            & $\{K^-,\Sigma^+\}$ & $2(D+F)\left(b_D+b_F\right)$ &
            $-2(D+F)\left(b_D+b_F\right)$ \\
      \hline
      $\Xi^-$ & $\{\pi^0,\Xi^-\}$ & 0 & $(D-F)\left(b_D-b_F\right)$ \\
            & $\{\pi^-,\Xi^0\}$ & $2(D-F)\left(b_D-b_F\right)$ & 0 \\
            & $\{\bar{K}^0,\Sigma^-\}$ & 0 & $2(D+F)\left(b_D+b_F\right)$ \\
            & $\{K^-,\Lambda \}$ & $\frac13(D-3 F)\left(b_D-3 b_F\right)$ & 0 \\
            & $\{K^-,\Sigma^0\}$ & $(D+F)\left(b_D+b_F\right)$ & 0 \\
            & $\{\eta_8,\Xi^-\}$ & 0 & $\frac13D+3 F)\left(b_D+3 b_F\right)$ \\
            & $\{\eta_0,\Xi^-\}$ & 0 & $\frac{2F_{\pi}^2}{3F_0^2} \beta$ \\
      \hline
   \end{tabular}
   \caption{\label{tab:CH} Possible loops contributing to
         the EDFFs of the hyperons and the corresponding
         coefficients $C_{\rm cd}$ and $C_{\rm ef}$. The intermediate states of the loops are listed in
         the second column for each hyperon.}
\end{center}
\end{table}

So far, there is no constraint on the LECs $w_{13,14}^{\prime\,r}(\mu)$ and
$w_{13,14}$, which appear at the LO of the baryon EDFFs, except for the $N_c$
scaling and the naturalness requirement of the effective field theory. As mentioned
in section~\ref{sec:EDFF}, the tree-level expression of the baryon EDFFs depend only
on two combinations
\begin{equation}
w_a(\mu) \equiv \alpha
w_{13}+w_{13}^{\prime\,r}(\mu)
\label{eq:w13}
\end{equation}
and $\alpha w_{14}+w_{14}^{\prime\,r}(\mu)$. In
principle, these two combinations may be extracted from lattice calculations of the
baryon EDMs. On the lattice, the calculations can be performed at different quark masses,
or equivalently pion masses. From a fitting to the pion mass dependence of some of
the baryon EDMs, one may extract the unknown LECs, and then make a prediction of the
other baryon EDMs. Furthermore, up to the order $\order{3}$,
\begin{equation}
   w_b(\mu) \equiv 3[\alpha w_{14}+w_{14}^{\prime\,r}(\mu)] + \frac{V_0^{(2)}
   \beta}{4\pi F_0^2F_\pi^2 m_{\rm ave}} M_{\eta_0}
\label{eq:w14beta}
\end{equation}
always appear together in the expressions of the baryon EDFFs because of SU(3)
flavor symmetry. The second term is due to the loops involving the $\eta_0$, and
its pion mass dependence starts from $\order{4}$. Here, all the baryon masses in
the second term have been replaced by the average mass of the baryons $m_{\rm ave}$
since the difference is a higher order effect. Consequently, NLO one-loop
expressions for all baryon EDFFs are given in terms of just {\em two} unknown LECs.

\subsection{Numerical results for the loop contributions of the baryon EDMs}
\label{sec:num}

\begin{figure}[th]
\centering
\includegraphics[width=\linewidth]{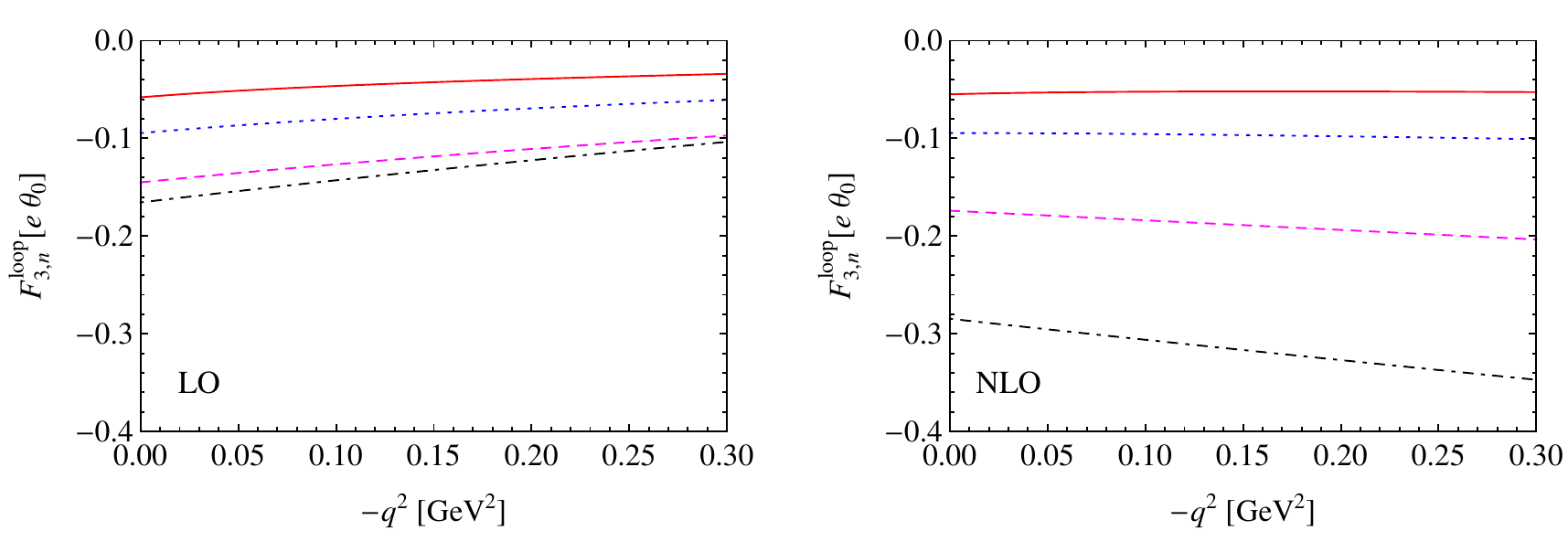}\hfill
\caption{Loop contribution to the neutron EDFF up to LO (left) and NLO (right).
The solid, dotted,
dashed and dot-dashed lines are for the pion mass 138~MeV (physical value),
200~MeV, 300~MeV and 400~MeV, respectively.
\label{fig:F3n}}
\end{figure}

\begin{figure}[th]
\centering
\includegraphics[width=0.49\linewidth]{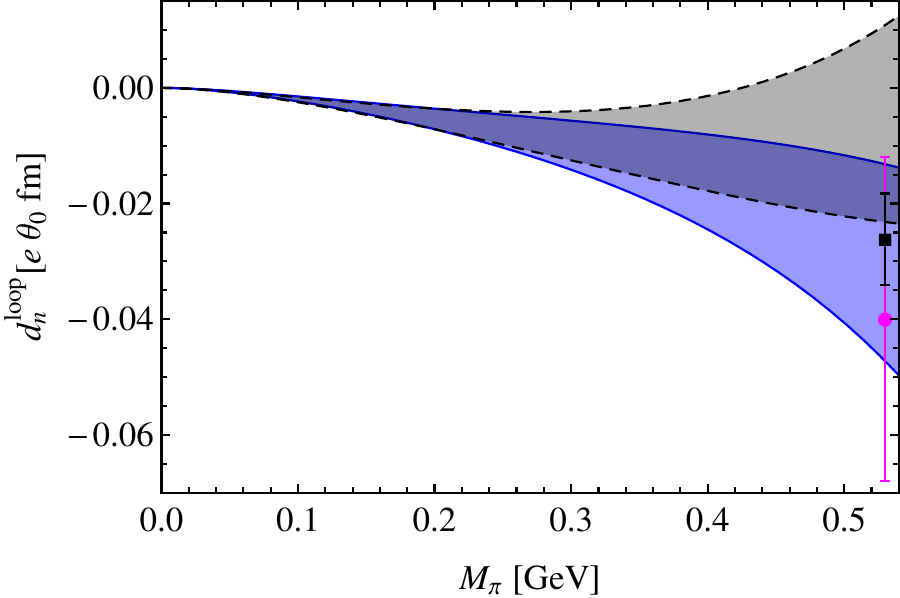}\hfill
\includegraphics[width=0.49\linewidth]{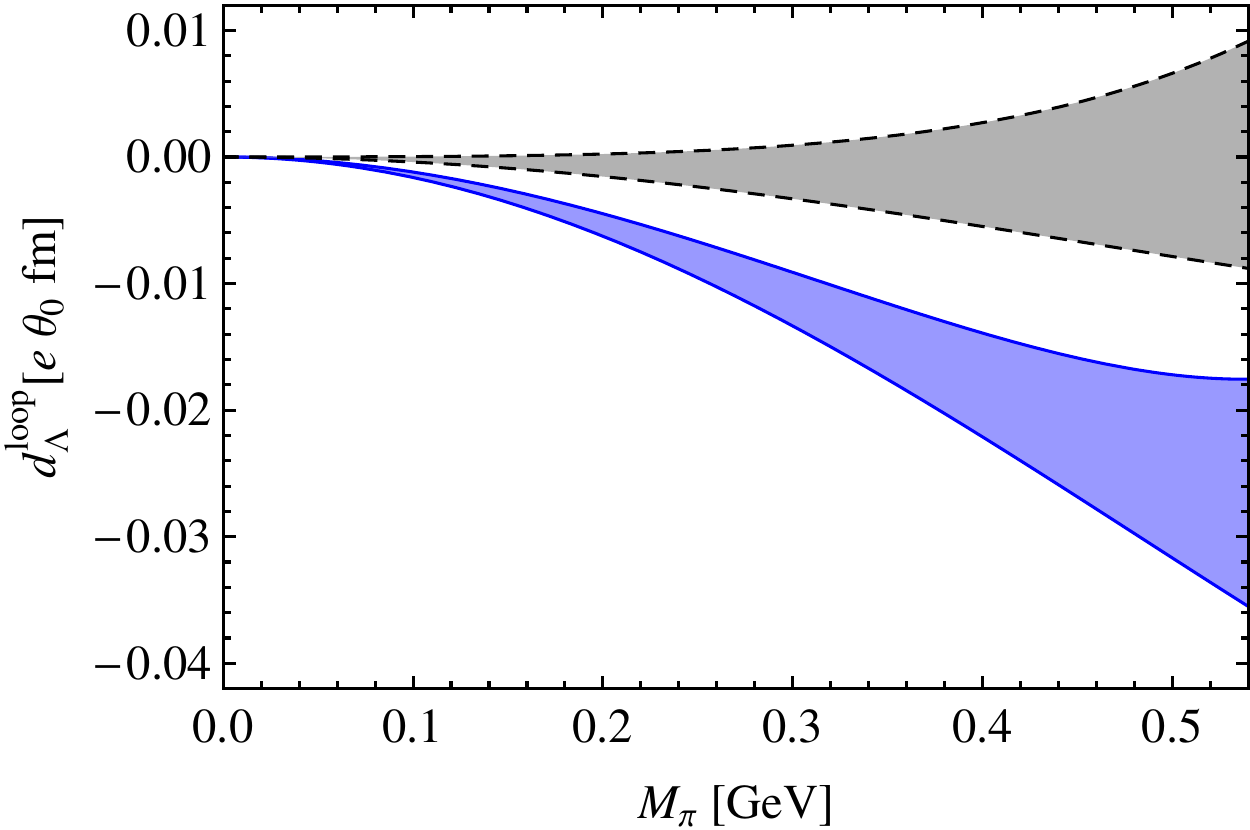}\\[4mm]
\includegraphics[width=0.49\linewidth]{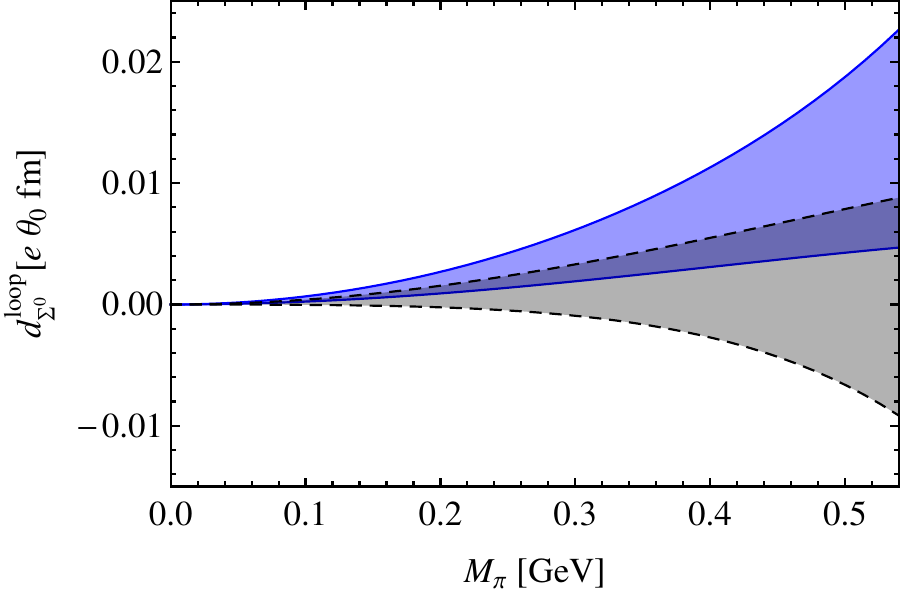}\hfill
\includegraphics[width=0.49\linewidth]{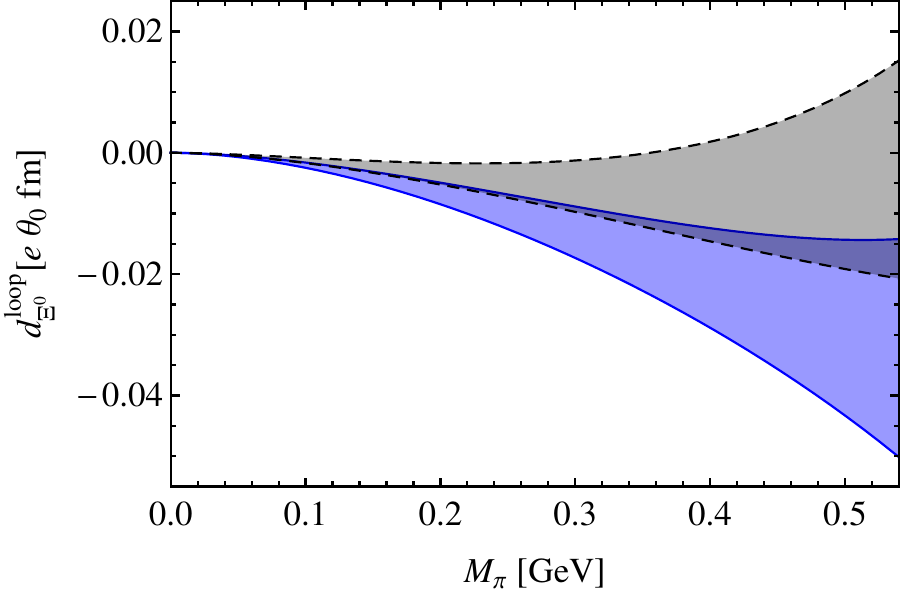}
\caption{Loop contributions to the EDMs of the neutral baryons as a function of the
pion mass.
The bands, reflecting uncertainties by varying the renormalization scale
between $\mu=M_\rho$ and $m_\Xi$, between solid and dashed boundaries are the NLO and the
LO results, respectively. The filled circle and square with error bars are the
lattice data from refs.~\cite{Shintani:2008nt} and
\cite{confX}, respectively.
\label{fig:dn}}
\end{figure}

Since the values of the LECs $w_{13,14}^{\prime\,r}(\mu)$ and $w_{13,14}$ are not
known, we will only focus on the contributions from the loops in this section, and
discuss several relations which are free of these parameters in
section~\ref{sec:relations}.
In order to get the numerical results, we use $D=0.804$ and $F=0.463$. From fitting
to the baryon mass differences at $\order{2}$, see appendix~\ref{app:mb}, we get
$b_D=0.068$~GeV$^{-1}$ and $b_F=-0.209$~GeV$^{-1}$. There are other determinations of
$b_D$ and $b_F$ from higher order analysis of various baryonic properties, see, for
instance, refs.~\cite{Borasoy:1996bx,Bsaisou:2012rg}. The difference reflects
higher order effects, so that they will not be used here. In the large
$N_c$-limit, one has $F_0=F_\pi$, and we take 92.2~MeV~\cite{PDG} for its value.
From an analysis of the $\eta-\eta'$
mixing in the framework of  U(3)$_L\times$U(3)$_R$ CHPT, it was found that
$V_0^{(2)}=-5\times10^{-4}$~GeV$^4$ and
$V_3^{(1)}=3.5\times10^{-4}$~GeV$^2$~\cite{HerreraSiklody:1997kd}. With these values,
the neutron EDFF calculated at $\mu=1$~GeV is plotted in figure~\ref{fig:F3n}, where
only the loop contribution is taken into account. One sees sizable NLO effects,
especially  for higher pion masses. One notices that the dependence on $q^2$ can be
well approximated by a linear function, which means that a linear extrapolation from
finite to vanishing $q^2$ can be used on the lattice. The same is true for the other
baryons, as none of them shows a strong $q^2$-dependence.

At the physical pion mass, we get the loop contributions to the baryon EDMs in
units of $10^{-16} e\,\theta_0$~cm,
\begin{eqnarray}
   d_n^{\rm loop} \al=\al -3.1\pm0.8, \qquad d_p^{\rm loop} = \phantom{+}5.6\pm1.0 -
   6.1\, (\beta\cdot \text{GeV}), \nonumber \\
   d_{\Lambda}^{\rm loop} \al=\al -2.6\pm0.4, \qquad d_{\Sigma^+}^{\rm loop} =
   \phantom{+}3.8\pm1.0 - 4.8\, (\beta\cdot \text{GeV}), \nonumber \\
   d_{\Sigma^0}^{\rm loop} \al=\al\phantom{+} 0.8\pm0.4, \qquad d_{\Sigma^-}^{\rm
   loop} = -2.1\pm0.2 + 4.8\, (\beta\cdot \text{GeV}), \nonumber \\
   d_{\Xi^0}^{\rm loop} \al=\al -3.6\pm0.8, \qquad d_{\Xi^-}^{\rm loop} =
   -3.7\pm0.2 + 4.3\, (\beta\cdot \text{GeV}),
   \label{eq:phyres}
\end{eqnarray}
where the uncertainties are estimated by varying the scale $\mu$ between the masses
of the $\rho$ and $\Xi$. If we replace the baryon masses in the $\beta$-term in the
baryon EDFF expressions, eqs.~\eqref{eq:EDFFp}, \eqref{eq:EDFFSp},
\eqref{eq:EDFFSm} and \eqref{eq:EDFFXm}, by the averaged baryon mass
$m_\text{ave}=1151$~MeV as that in eq.~\eqref{eq:w14beta}, this term contributes
$-5.0\,(\beta\cdot \text{GeV})$ to the proton and $\Sigma^+$ and $5.0\,(\beta\cdot
\text{GeV})$ to the $\Sigma^-$ and $\Xi^-$. The difference reflects part of the
higher order uncertainties. In the following, we will use the averaged baryon mass
for the $\beta$-term and keep different masses for the other terms.

In order to compare with the results from lattice simulations, we should study the
pion mass dependence of the pertinent quantities. We take the physical values for
$F_\pi$, $M_{\eta_8,\eta_0}$ and the baryon masses, since their $M_\pi$-dependent
effects contribute at higher orders. For the kaon mass, we use
\begin{equation}
   M_K^2 = \mathring{M}_K^2 + \frac{M_\pi^2}2,
\end{equation}
where $\mathring{M}_K=484$~MeV is the kaon mass in the SU(2) chiral limit with
vanishing up and down quark masses. The pion mass dependence of the EDMs of the
neutral baryons is shown in figure~\ref{fig:dn}, where the shaded bands
between the solid
and dashed boundaries are the NLO and LO results, respectively. It is clear that all
the EDMs vanishes in the chiral limit. The physical reason is that the $\theta$-term
can be rotated away if any of the quarks are massless, see, for instance,
ref.~\cite{Donoghue}, which ensures a vanishing value for the $\theta$-term induced
EDM.

For comparison, the lattice data for the neutron at $M_\pi=530$~MeV
calculated in refs.~\cite{Shintani:2008nt} are shown, which are calculated on
$24^3\times48$ lattice with lattice spacing $a\approx0.11$~fm. We also show the lattice
data reported very recently in ref.~\cite{confX} whose uncertainty is smaller.
Both the LO and NLO loop results agree with the lattice data.

\begin{figure}[t]
\centering
\includegraphics[width=0.49\linewidth]{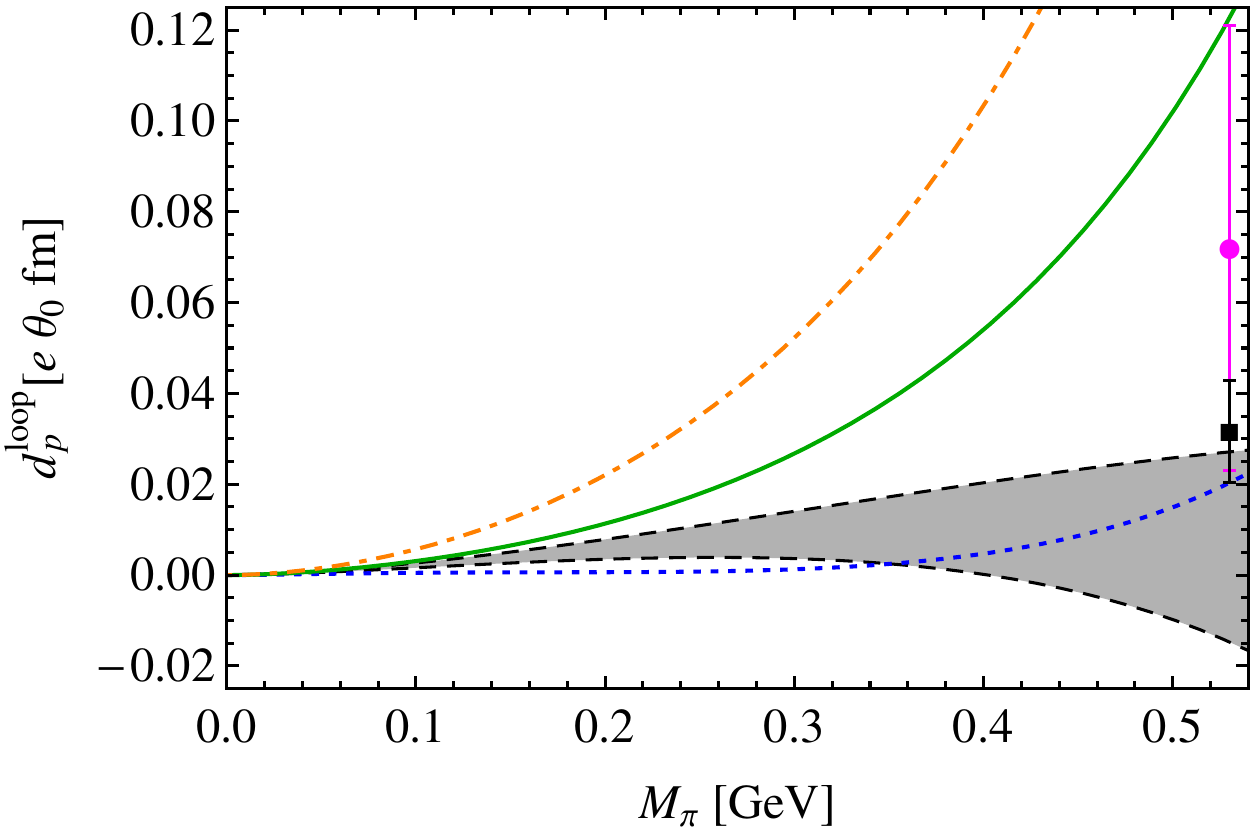}\hfill
\includegraphics[width=0.49\linewidth]{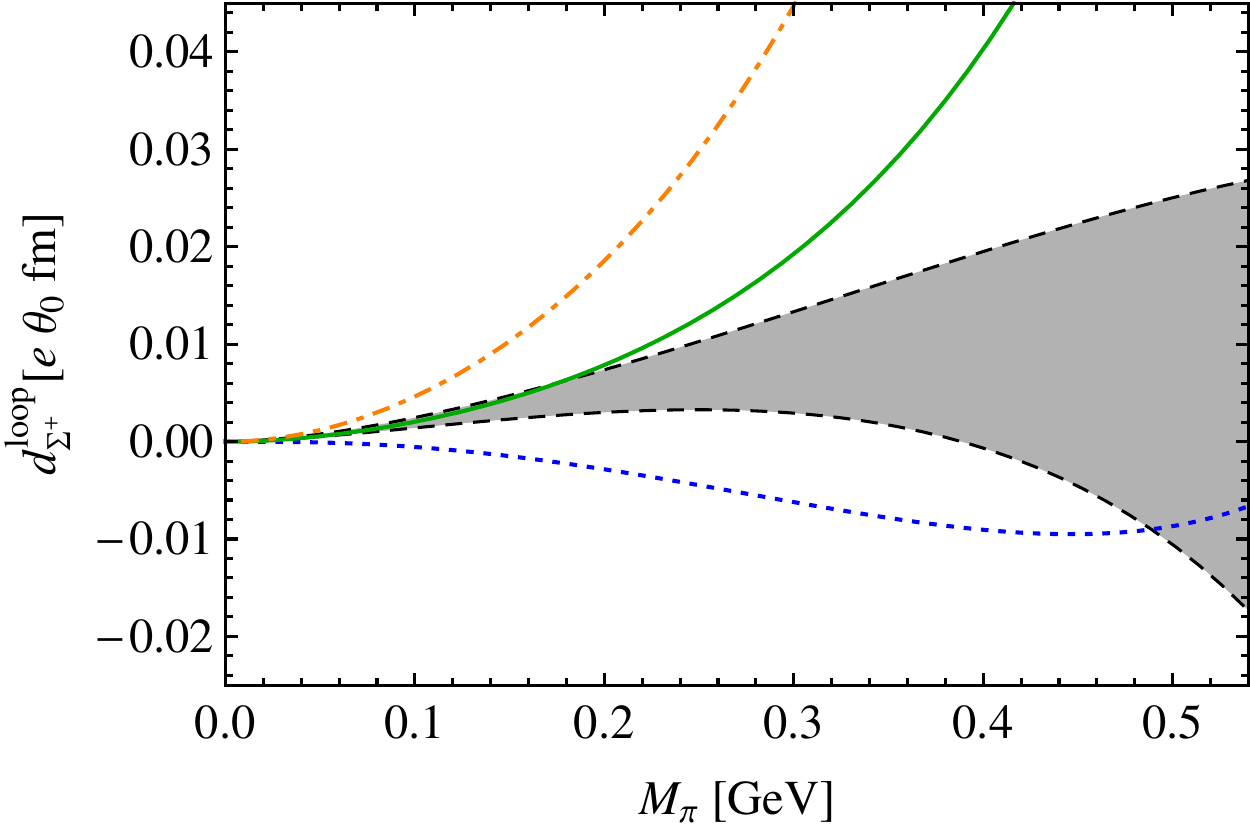}\\[4mm]
\includegraphics[width=0.49\linewidth]{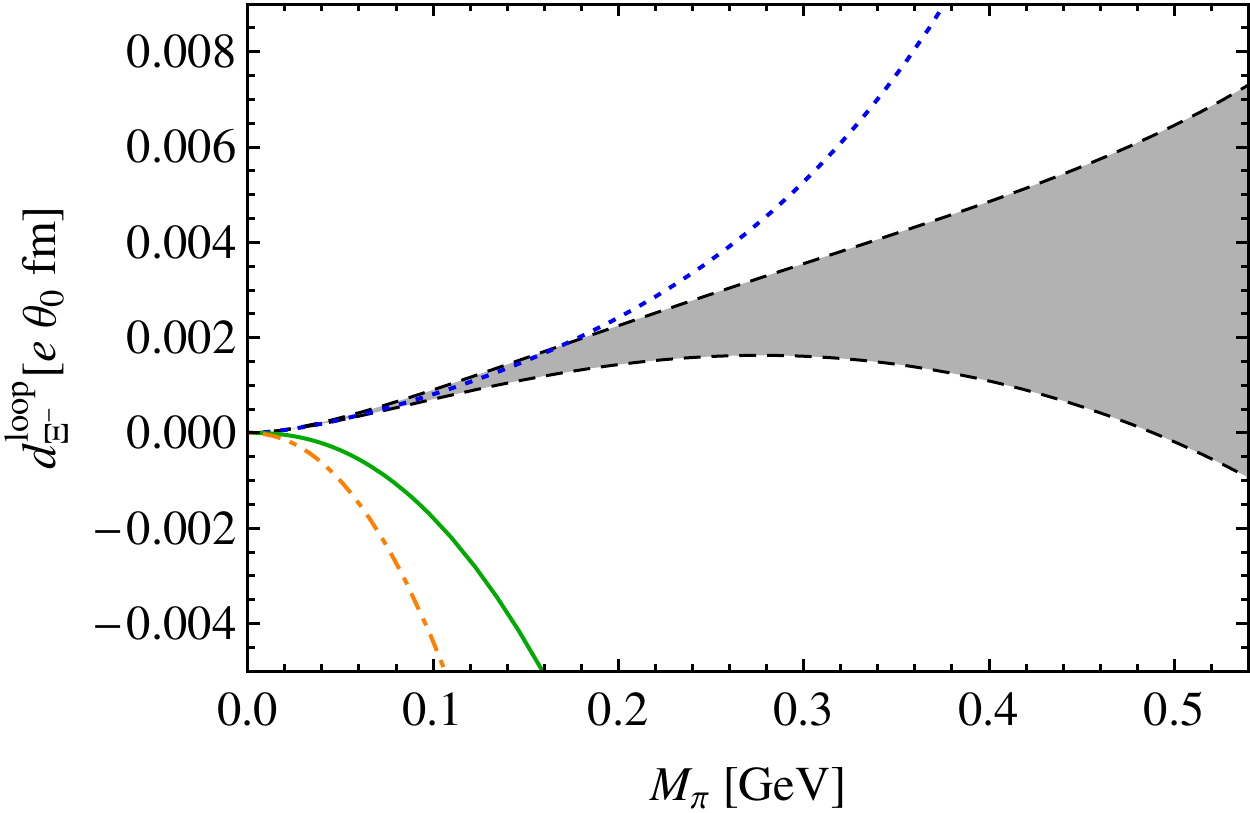}\hfill
\includegraphics[width=0.49\linewidth]{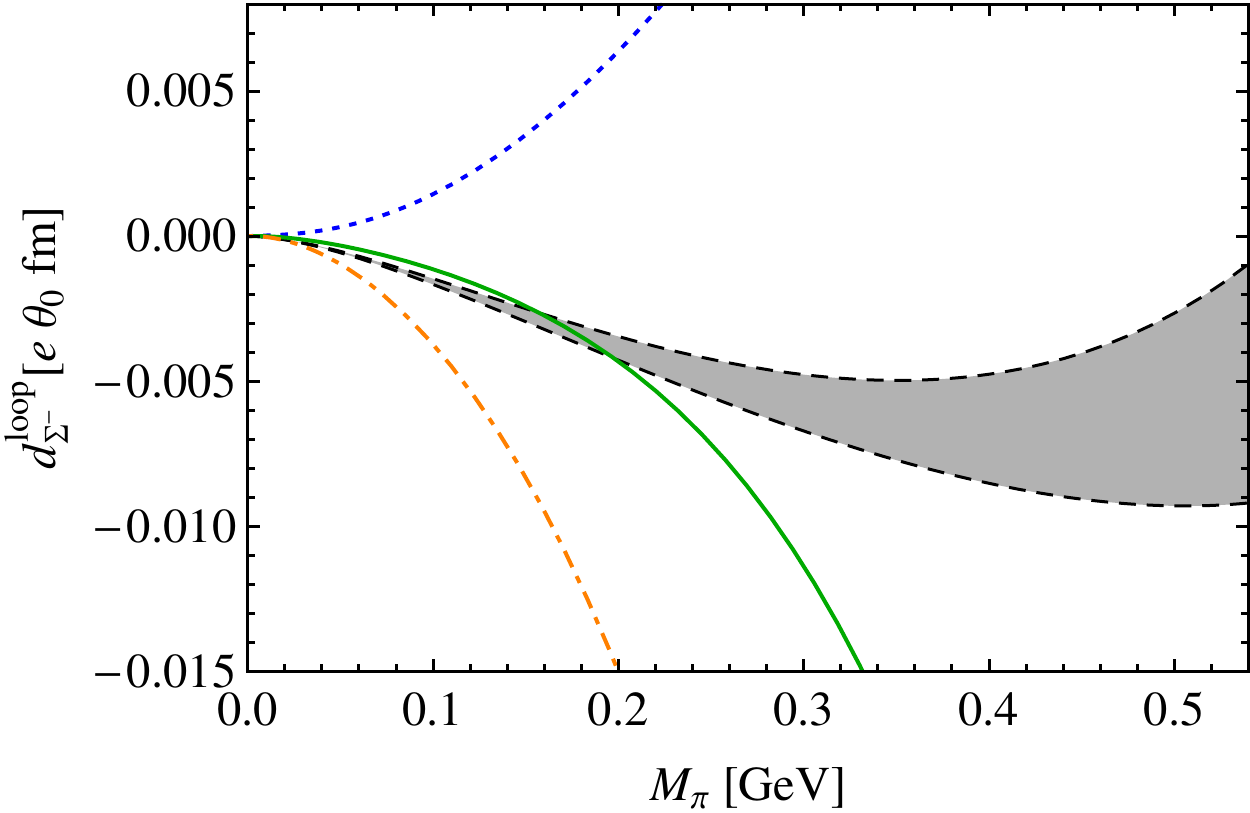}
\caption{Loop contributions to the EDMs of the charged baryons as a function of the
pion mass.
The bands, reflecting uncertainties of the LO contributions by varying the
renormalization scale between $\mu=m_\rho$ and $m_\Xi$. The filled circle and square
with error bars are the lattice data from refs.~\cite{Shintani:2008nt} and
\cite{confX}, respectively.
The solid, dotted and dot-dashed lines represent the NLO results evaluated at
$\mu=1$~GeV with $\beta = 0,1$ and $-1$~GeV$^{-1}$, respectively.
\label{fig:dc}}
\end{figure}
One notices that the NLO corrections for all the neutral hyperons are dramatic. For
the $\Lambda$, from table~\ref{tab:CN} and eq.~\eqref{eq:F3loop}, one finds that
the loops involving $\pi^+$ and $\Sigma^-$ cancel with those of $\pi^-$ and
$\Sigma^+$ exactly. Thus, the remaining LO contributions come from loops involving a
kaon. They are small as can be seen from
$$ 1+ \ln \frac{M_\pi^2}{\mu^2} = -2.96, \qquad 1+ \ln \frac{M_K^2}{\mu^2} =
-0.40~, $$
with $\mu=1$~GeV.  Therefore, the LO result for the $\Lambda$ EDM is close to
zero (as already pointed out in ref.~\cite{Borasoy:2000pq}). The same happens
for the $\Sigma^0$.

The results for the charged baryons are shown in figure~\ref{fig:dc}, where the bands
are the LO loop results. We choose three different values for the unknown
combination $\beta$ to illustrate the NLO effects. The results are quite sensitive
to the numerical value of $\beta$, especially for the $\Xi^-$ and $\Sigma^-$.

\subsection{A first determination of the  LECs from lattice results}

We may  use the neutron and the proton EDMs at unphysical quark masses
to determine the two combination of
parameters $w_a(\mu)$ and $w_b(\mu)$. Using the lattice data at $M_\pi =
530\,$MeV~\cite{confX}\footnote{As this pion mass value
  is at the edge of the range of applicability of our approach, lattice data
  at lower pion mass are urgently called for for a more reliable determination of the LECs.}
for the neutron EDM, one may determine the counterterm combination $w_a(\mu)$
\begin{equation}
   w_a(1~\text{GeV}) = (-0.01\pm0.02)~\text{GeV}^{-1},
    \label{eq:Nwa}
\end{equation}
where the uncertainty merely reflects the uncertainty in the lattice calculations.
Using the data for the proton, we get
\begin{equation}
   w_b(1~\text{GeV}) = (-0.40\pm0.05)~\text{GeV}^{-1}.
   \label{eq:Nwb}
\end{equation}

With these determinations, the EDMs at the physical pion mass can be predicted
including both the tree and the loop contributions.
The ones for the neutron and proton in units of $10^{-16} e\,\theta_0\,$cm are
\begin{equation}
   d_n = -2.9\pm0.4\pm0.8, \qquad d_p = 1.1\pm0.5\pm1.0.
\end{equation}
Here, the first uncertainty reflects the uncertainty in the determination of
$w_a(1~\text{GeV})$ and $w_b(1~\text{GeV})$, and the second one corresponds to varying
the scale $\mu$ between the $\rho$-meson mass and $m_\Xi$. Comparing with the loop
results given in eq.~\eqref{eq:phyres}, one sees that the neutron EDM at the physical
pion mass is dominated by the loops. For the proton, the loop contributions have
similar size as, albeit slightly larger than, the tree level terms.
Combining these two errors, we have  $d_n = -2.9\pm 0.9$ and $d_p = d_p =
1.1\pm 1.1$ (in canonical units). Using the
experimental upper limit of the neutron EDM,
$|d_n|<2.9\times10^{-26}e$~cm~\cite{Baker:2006ts}, the $\theta_0$ angle is
constrained to be
\begin{equation}
   |\theta_0|\lesssim 1.5\times10^{-10}~.
\end{equation}
Similarly, we can predict the hyperon EDMs (again in units of $10^{-16}
e\,\theta_0$~cm)
\begin{eqnarray}
   d_\Lambda \al=\al -2.5\pm0.2\pm0.4, \qquad d_{\Sigma^+} = -0.7\pm0.5\pm1.0,
   \nonumber\\
   d_{\Sigma^0} \al=\al \phantom{+}0.7\pm0.2\pm0.4,\qquad  d_{\Sigma^-} =
   \phantom{+}2.2\pm0.5\pm0.2 \nonumber\\
   d_{\Xi^0} \al=\al -3.4\pm0.4\pm0.8,\qquad  d_{\Xi^-} = \phantom{+}0.6\pm0.5\pm0.2.
\end{eqnarray}

\subsection{Counterterm-free relations}
\label{sec:relations}

\begin{table}[t]
\begin{center}
   \begin{tabular}{|l | c c c c c c c c |} \hline
      Baryon      & $p$      & $n$  & $\Sigma^+$ & $\Sigma^0$ & $\Sigma^-$ &
      $\Lambda$ & $\Xi^0$ & $\Xi^-$
      \\
      \hline
      Combination & $-(w_a+w_b)$ & $2w_a$ & $-(w_a+w_b)$ & $-w_a$ & $w_b-w_a$ &
      $w_a$     & $2w_a$  & $w_b-w_a$
      \\
      \hline
   \end{tabular}
   \caption{\label{tab:unknown} Combinations of the unknown parameters
     appearing in the EDFFs of the baryons up to NLO.}
\end{center}
\end{table}

In this section, we will derive relations that are free of the unknown LECs up to
NLO. These relations can e.g. serve as checks of lattice simulations for varying
quark masses. In fact, due the  SU(3) flavor symmetry, there exists an exact
relation among the EDFFs of the $\Sigma$ hyperons. From table~\ref{tab:CN}, one
deduces that the loops involving a pion or an $\eta_{8(0)}$ for the $\Sigma^+$
cancel with those of the $\Sigma^-$. As mentioned before, the pionic loops of the
$\Sigma^0$ cancel with each other. Furthermore, there are also cancellations for
the kaonic loops among the $\Sigma$ hyperons. These cancellations occur when
isospin is a good symmetry. Therefore, in the isospin symmetric case, which is
considered throughout, one has
\begin{equation}
   F_{3,\Sigma^+} + F_{3,\Sigma^-} - 2 F_{3,\Sigma^0} = \order{4}.
   \label{eq:relation1}
\end{equation}
This relation is exact up to NLO including the tree-level contributions. Similarly,
there is another relation which is exact only at LO
\begin{equation}
   F_{3,\Sigma^0} + F_{3,\Lambda} = \order{3},
   \label{eq:F3com1}
\end{equation}
since the LO loop contributions are independent of the baryon mass, and they cancel
each other in the sum. Moreover, the above combination is also independent of $\beta$
which combines the unknown parameters $w_0$ and $w_{10}'$, so that it can be
calculated parameter-free. Taking $q^2=0$, the $\order{3}$ expression is rather
simple
\begin{eqnarray}
   d_{\Sigma^0} + d_{\Lambda} = -\frac{16 e V_0^{(2)} \bar{\theta}_0}{3
   \pi F_\pi^4 M_K} \left(M_K^2-M_\pi^2\right) \left(F b_D^2+2 D b_D b_F+3 F
   b_F^2\right) + \order{4}.
   \label{eq:relation2}
\end{eqnarray}

\begin{figure}[t]
\centering
\includegraphics[width=\linewidth]{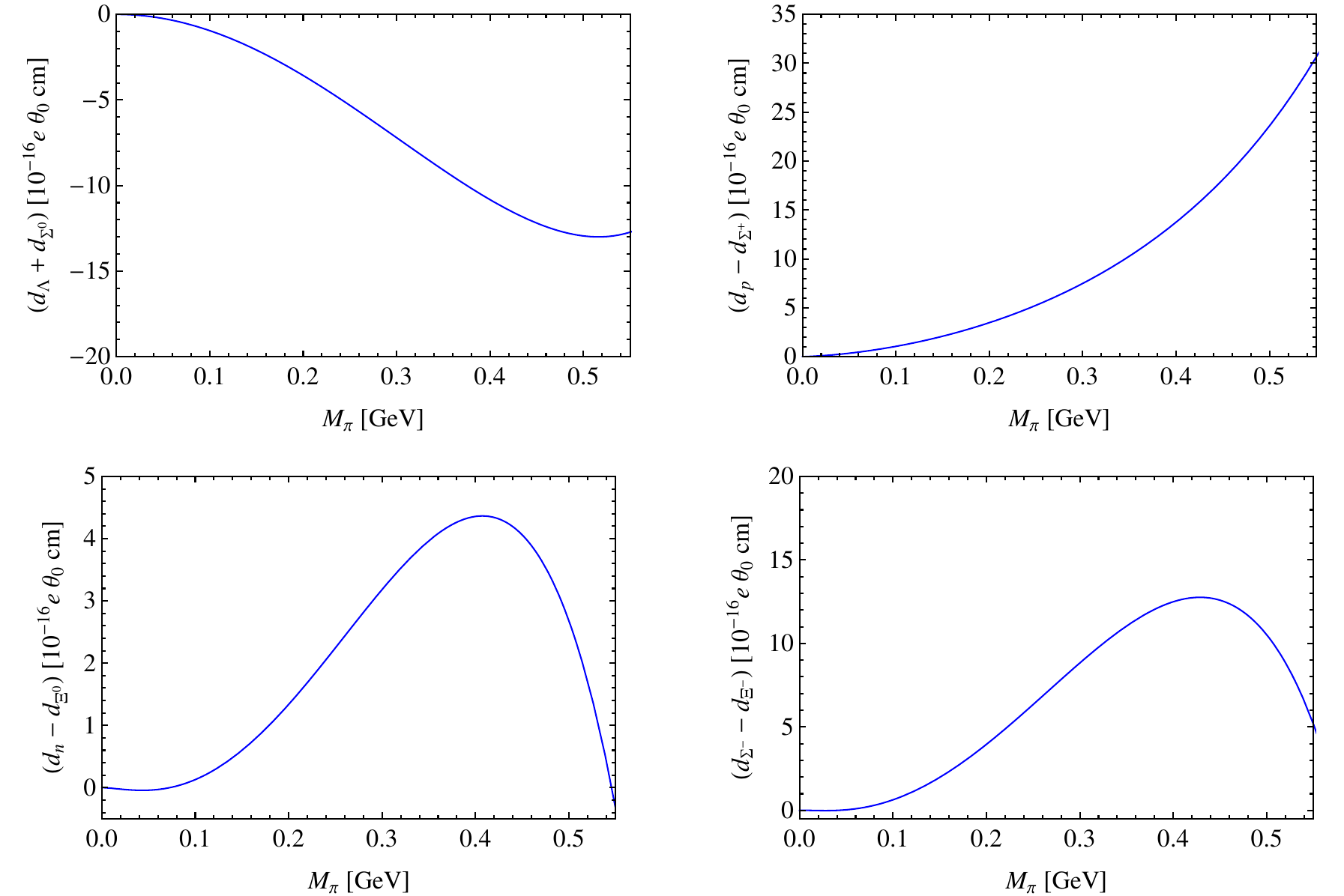}
\caption{Counterterm-free combinations of the baryon EDMs as a function of the pion
mass.
\label{fig:relations}}
\end{figure}

In fact, there are more combinations free of the counterterms. As mentioned
before, up to NLO, there are only two combinations of unknown constants,
$w_a(\mu)$ and $w_b(\mu)$, in the expressions  of the EDFFs of the octet
baryons defined in eqs.~\eqref{eq:w13} and \eqref{eq:w14beta}.
It is easy to find combinations free of any unknown parameters up to
$\order{3}$ utilizing the results collected in table~\ref{tab:unknown}.
In addition to the two in eqs.~\eqref{eq:relation1} and
\eqref{eq:relation2}, they include
\begin{eqnarray}
   F_{3,p} - F_{3,\Sigma^+}, \quad F_{3,n} - F_{3,\Xi^0}, \quad
   F_{3,\Sigma^-} - F_{3,\Xi^-}.
   \label{eq:F3com}
\end{eqnarray}
Because the counterterms have been cancelled out, these combinations are also
finite and hence independent of  the scale $\mu$. However, different from
eq.~\eqref{eq:relation2}, their LO loop contributions do not vanish. The expression
for the $F_{3,n} - F_{3,\Xi^0}$ at $q^2=0$ reads
\begin{eqnarray}
   d_{n} - d_{\Xi^0} \al=\al \frac{e V_0^{(2)} \bar{\theta}_0}{\pi^2 F_\pi^4}
   \bigg[ \left( D b_D + F b_F \right) \left( 2 \ln \frac{M_K^2}{M_\pi^2} +
   \pi \frac{M_\pi - M_K}{m_{\rm ave}} \right) \nonumber \\
   \al\al+ \frac{8\pi}{M_K}
   \left(M_K^2-M_\pi^2\right) \left(D b_D^2+2 F b_D b_F+D b_F^2\right) \bigg] +
   \order{4}.
   \label{eq:relation3}
\end{eqnarray}
The relations for the charged baryons are more complicated, and are not shown here
(they can be obtained from the expressions given in appendix~\ref{app:expressions}).
The pion mass dependence of the four combinations in eqs.~\eqref{eq:F3com1} and
\eqref{eq:F3com} are shown in figure~\ref{fig:relations}. Their values at the
physical pion mass in units of $10^{-16}\,e\,\theta_0$~cm are
\begin{eqnarray}
   d_{\Sigma^0} + d_{\Lambda} \al=\al -1.8,\qquad  d_{p} - d_{\Sigma^+} \,\,\, =
   1.8~,\nonumber\\
   d_{n} - d_{\Xi^0} \al=\al \phantom{-}0.5, \qquad d_{\Sigma^-} - d_{\Xi^-} = 1.6~.
\end{eqnarray}
Note that although the first combination starts from one order higher than the
others, their numerical values are of similar order of magnitude.

\section{Finite-volume corrections in the \texorpdfstring{$p$}{p}-regime}
\label{sec:fv}

On the lattice, calculations are performed in a finite volume.
As a result, the continuum momentum spectrum becomes quantized. Taking  periodic
boundary condition for all three spatial dimensions forms a torus. If the
volume is $L^3$, the momentum takes values of $2\pi \vec{n}/L$, with
$\vec{n}$ a three-dimensional vector
of integers. One easily sees that any integral over the spatial components of
momentum in the infinite volume should be replaced by a summation for the
momentum modes. For instance, the
two-point scalar loop integral needs to be changed as follows,
\begin{equation}
   i \int \frac{d^4 k}{(2\pi)^4} \frac1{(k^2-m_1^2)[(k+q)^2-m_2^2]} \rightarrow
   \frac{i}{L^3}
   \sum_{\vec{n}} \int \frac{d k^0}{2\pi} \frac1{(k^2-m_1^2)[(k+q)^2-m_2^2]}.
\end{equation}

The finite volume corrections to a quantity $\mathcal{Q}$ is defined as the
difference of $\mathcal{Q}$ evaluated in a finite and infinite volume,
\begin{equation}
   \delta_L[\mathcal{Q}] = \mathcal{Q}(L) - \mathcal{Q}(\infty).
\end{equation}

As we already mentioned, at chiral limit the baryon EDMs should vanish. However,
the current lattice data at rather large pion masses do not show a decreasing
behavior yet~\cite{Shintani:2006xr,Shintani:2008nt}. Thus, results at smaller pion
masses are necessary. However, in the section, we will show that the finite volume
corrections at small pion masses to most of the baryon EDMs are quite large, and it
is necessary to include the NLO contributions. The decomposition of the matrix
element of the electromagnetic current in the form of eq.~\eqref{eq:formfactors}
assumes Lorentz invariance. However, on discretized lattice, Lorentz invariance in
infinite volume is lost. Furthermore, the external momentum is also quantized in a
finite volume. These effects should be taken into account when calculating the
finite volume corrections to the form factors~\cite{Tiburzi:2007ep,Greil:2011aa}.
Since we are not aiming at a very precise calculation of the finite volume effects,
we will simply assume Lorentz invariance in the following.

As shown in ref.~\cite{Gasser:1987zq}, one can use the same LECs as in the infinite
volume, and the finite volume corrections come only from loops. This may be
understood easily, since the finite volume effects are long-distance physics, while
the LECs reflect the short-distance physics. For the same reason, the finite volume
corrections should not depend on the choice of method to regularize the ultraviolet
divergence. This means that one may change the upper bound of the Feynman parameter
integration in eqs.~\eqref{eq:DeltaLJMMm} and \eqref{eq:DeltaLJMmm} in the infrared
regularization from $\infty$ to $1$ as already noticed in, for instance,
ref.~\cite{Geng:2011wq}. In the $p$-regime, where $M_\pi L\gg
1$~\cite{Gasser:1986vb,Gasser:1987zq}, the finite volume corrections are
exponentially suppressed.

\begin{figure}[t]
\centering
\includegraphics[width=0.5\linewidth]{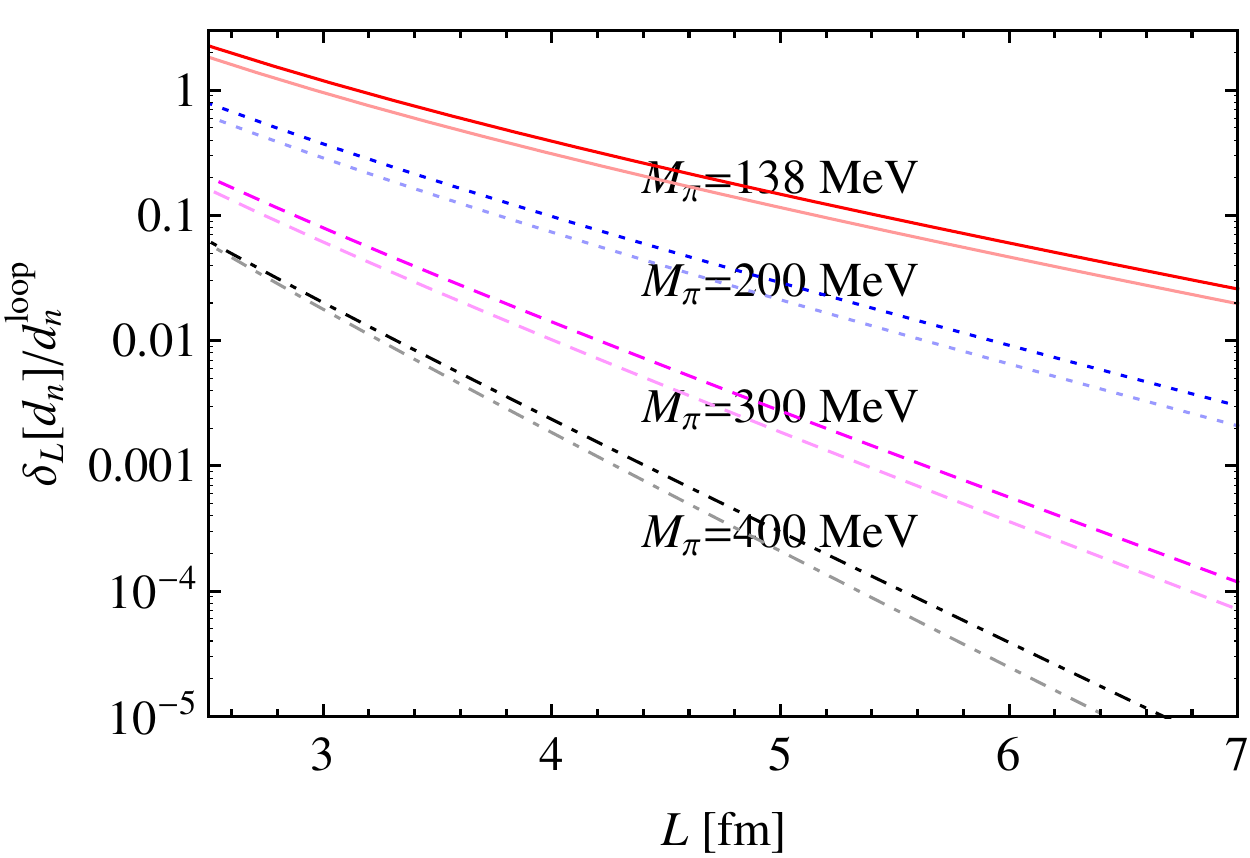}
\hfill\includegraphics[width=0.48\linewidth]{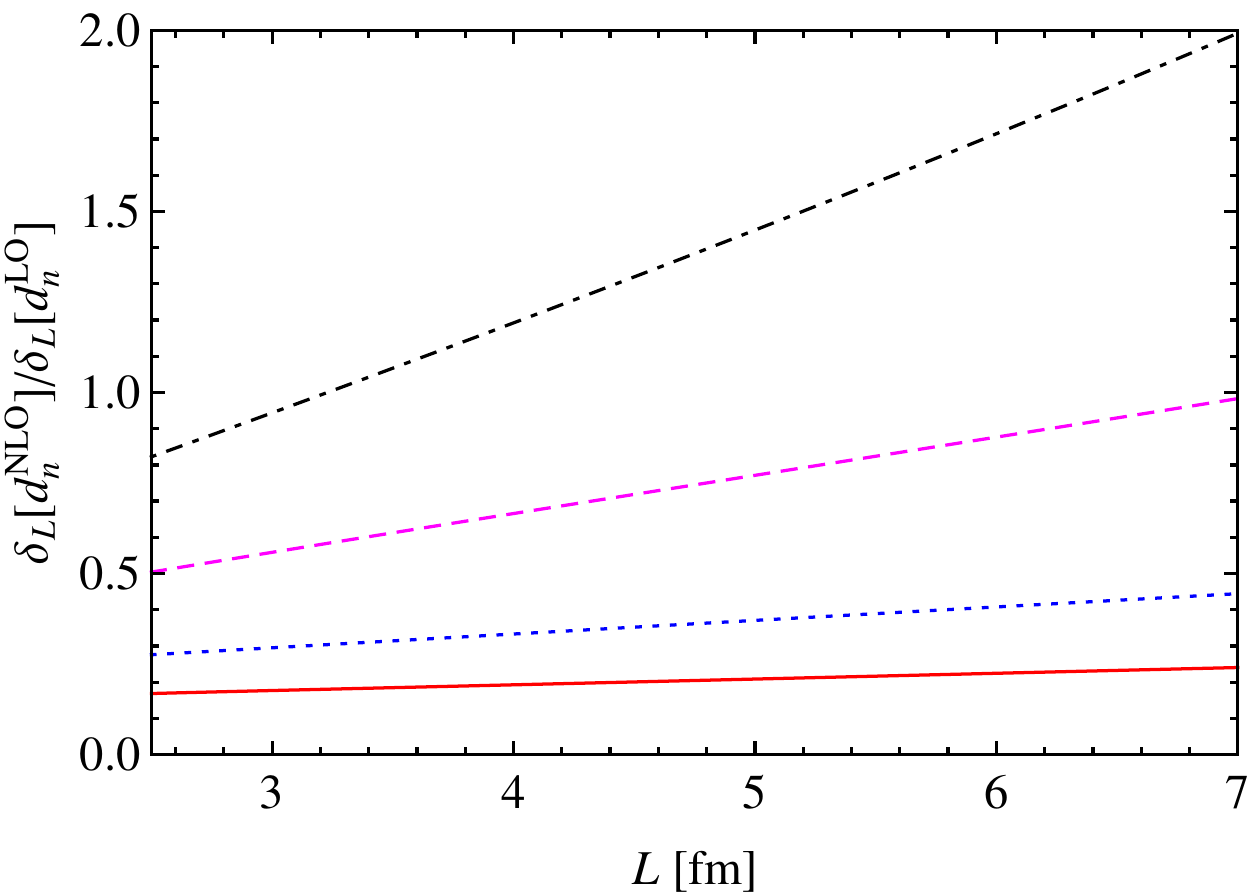}
\caption{Left: The ratios of the finite volume corrections to the loop contributions
in the infinite volume for the neutron EDM as a function of $L$. Right: The ratios of the
finite volume corrections at NLO (LO not included) to those at LO. The solid, dotted,
dashed and dot-dashed lines are for the pion mass being physical, 200~MeV, 300~MeV and 400~MeV,
respectively. In the left panel, for each value of the pion mass, two lines are
plotted with the upper and lower ones representing the NLO and LO results, respectively.
\label{fig:nEDMFV}}
\end{figure}
In ref.~\cite{O'Connell:2005un}, finite volume correction to the neutron EDM was
calculated at LO in SU(2) heavy baryon CHPT. The resulting expression is
given by~\cite{O'Connell:2005un}
\begin{equation}
   \delta_L\! \left[ d_n^{\rm LO} \right] =
   \frac{g_A \bar \alpha e \theta_0}{2\pi^2 F_\pi^2} \frac{m_u m_d}{m_u+m_d}
    \sum_{\vec{n}\neq0} K_0\left( L M_\pi |\vec{n}| \right),
    \label{eq:ConnellSavage}
\end{equation}
where $\bar\alpha$ is a coefficient of the CP-violating $NN\pi$
vertex.~\footnote{$\bar\alpha$ is the $\alpha$ used in
ref.~\cite{O'Connell:2005un}.}
From eqs.~\eqref{eq:F3n} and \eqref{eq:DeltaLJm1m2}, our result for the finite volume
correction to the neutron EDM at LO is
\begin{equation}
   \delta_L\! \left[ d_n^{\rm LO} \right] =
   \frac{M_\pi^2 e \theta_0}{4\pi^2 F_\pi^2}
    \sum_{\vec{n}\neq0} \big[ (D+F)\left(b_D+b_F\right) K_0\left( L M_\pi
   |\vec{n}| \right) - (D-F)\left(b_D-b_F\right) K_0\left( L M_K
   |\vec{n}| \right) \big]~,
\end{equation}
where we have used~\cite{Borasoy:2000pq}
\begin{equation}
   \bar\theta_0 \simeq \frac{F_\pi^2 M_\pi^2}{8 V_0^{(2)}} \theta_0,
\end{equation}
which may be obtained
from eq.~\eqref{eq:theta0barMK} considering $M_\pi\ll M_K$. Comparing the SU(2) part
with eq.~\eqref{eq:ConnellSavage}, we can identify
$$ \bar\alpha = \frac{B_0}2 ( b_D+b_F). $$

\begin{figure}[t]
\centering
\includegraphics[width=0.5\linewidth]{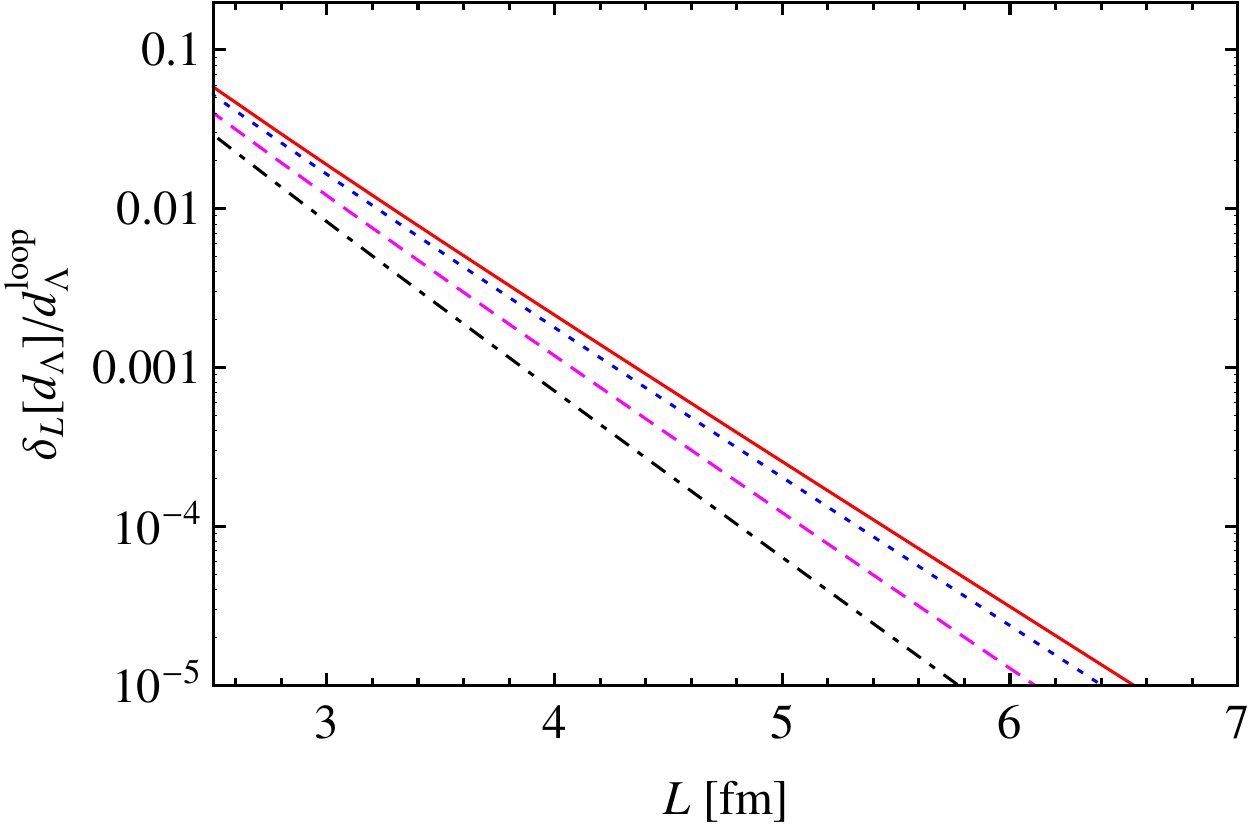}
\hfill\includegraphics[width=0.48\linewidth]{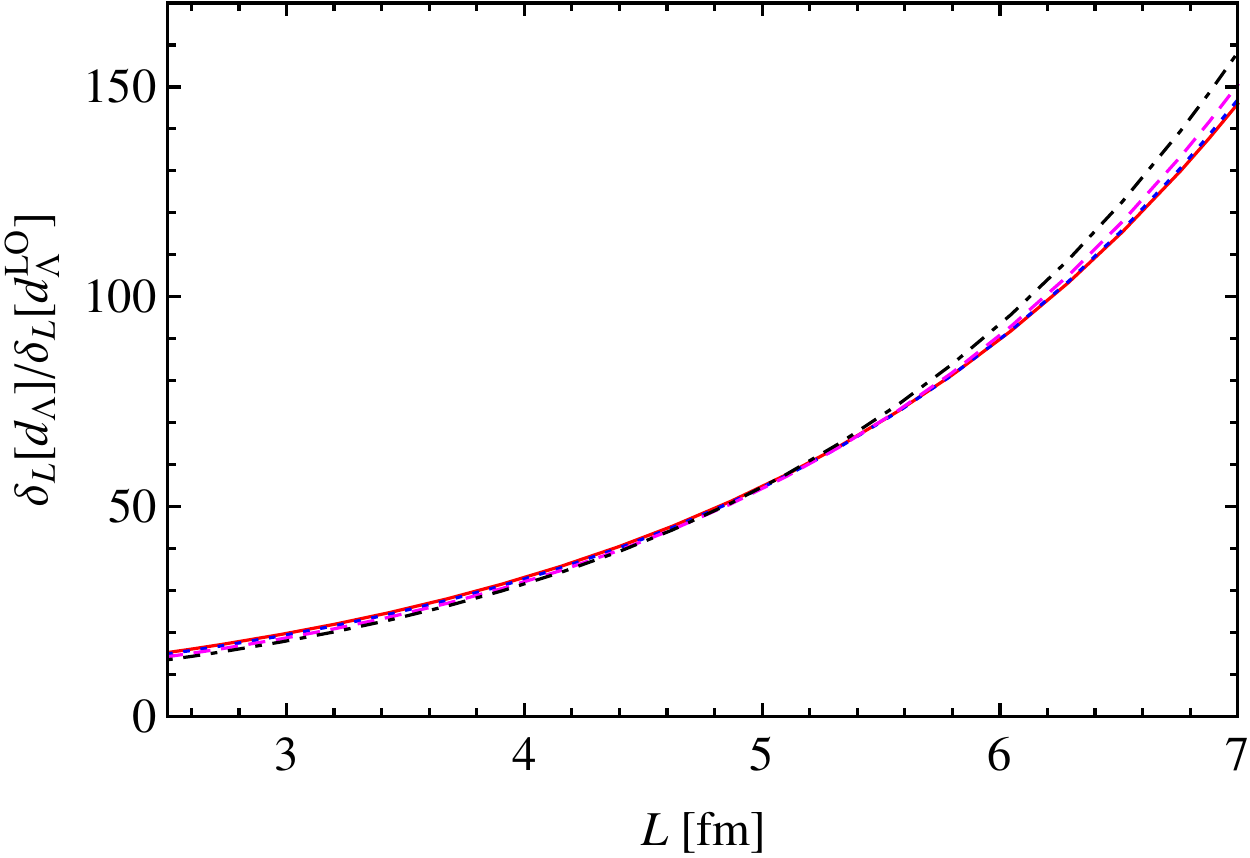}\\[2mm]
\includegraphics[width=0.5\linewidth]{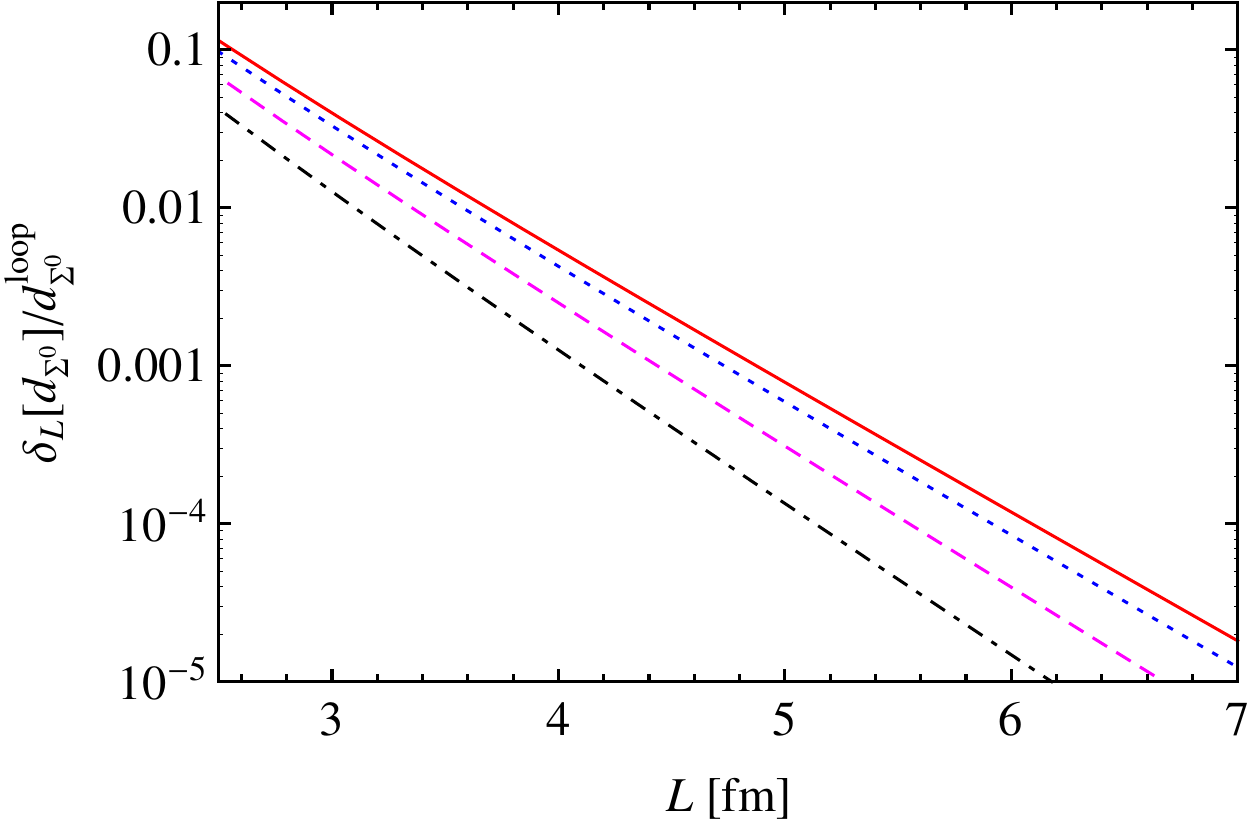}
\hfill\includegraphics[width=0.48\linewidth]{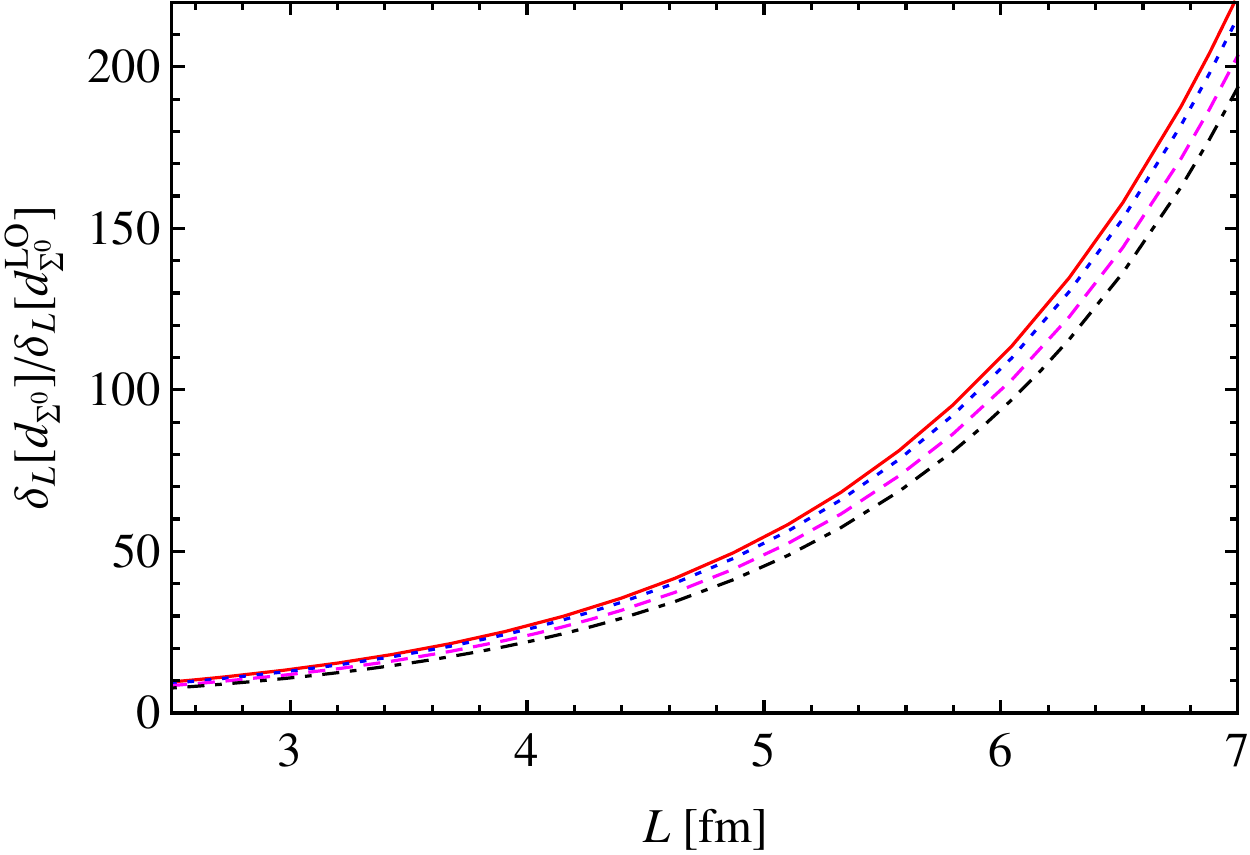}\\[4mm]
\includegraphics[width=0.5\linewidth]{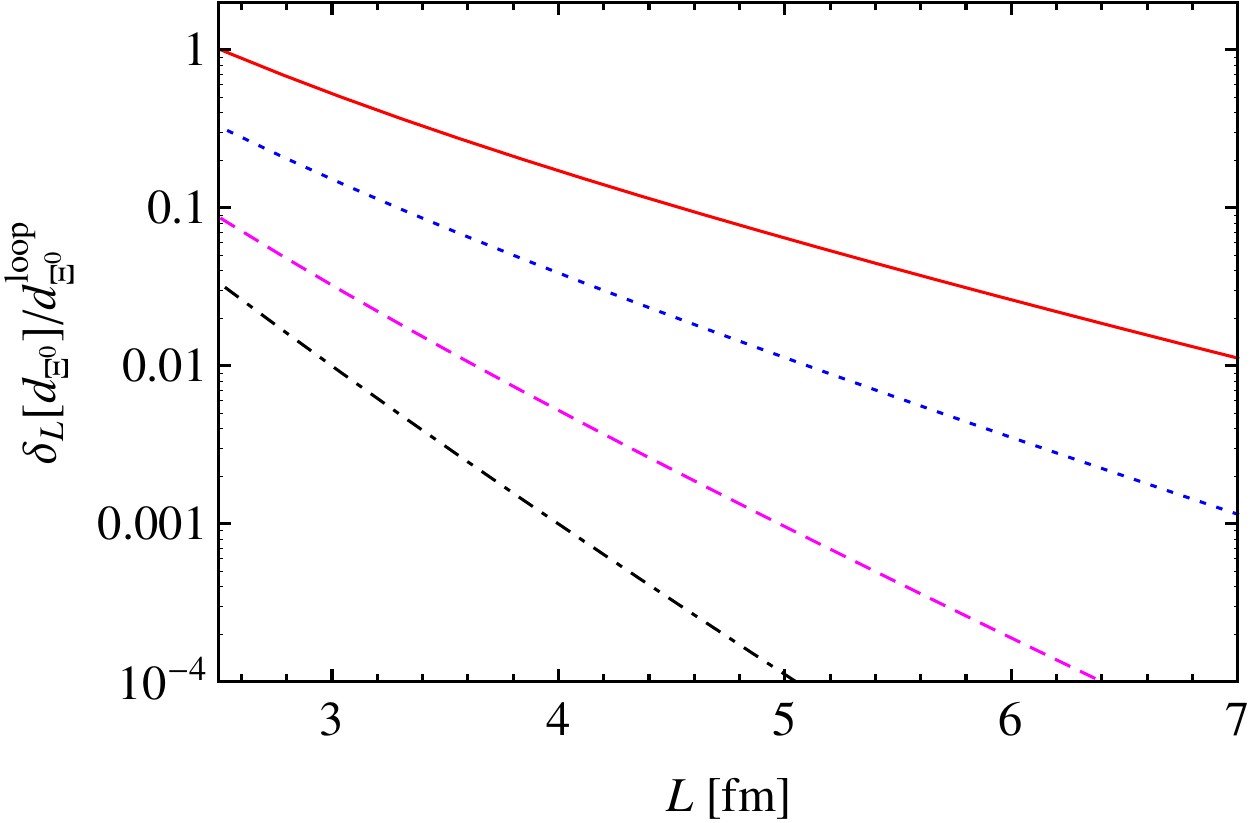}
\hfill\includegraphics[width=0.48\linewidth]{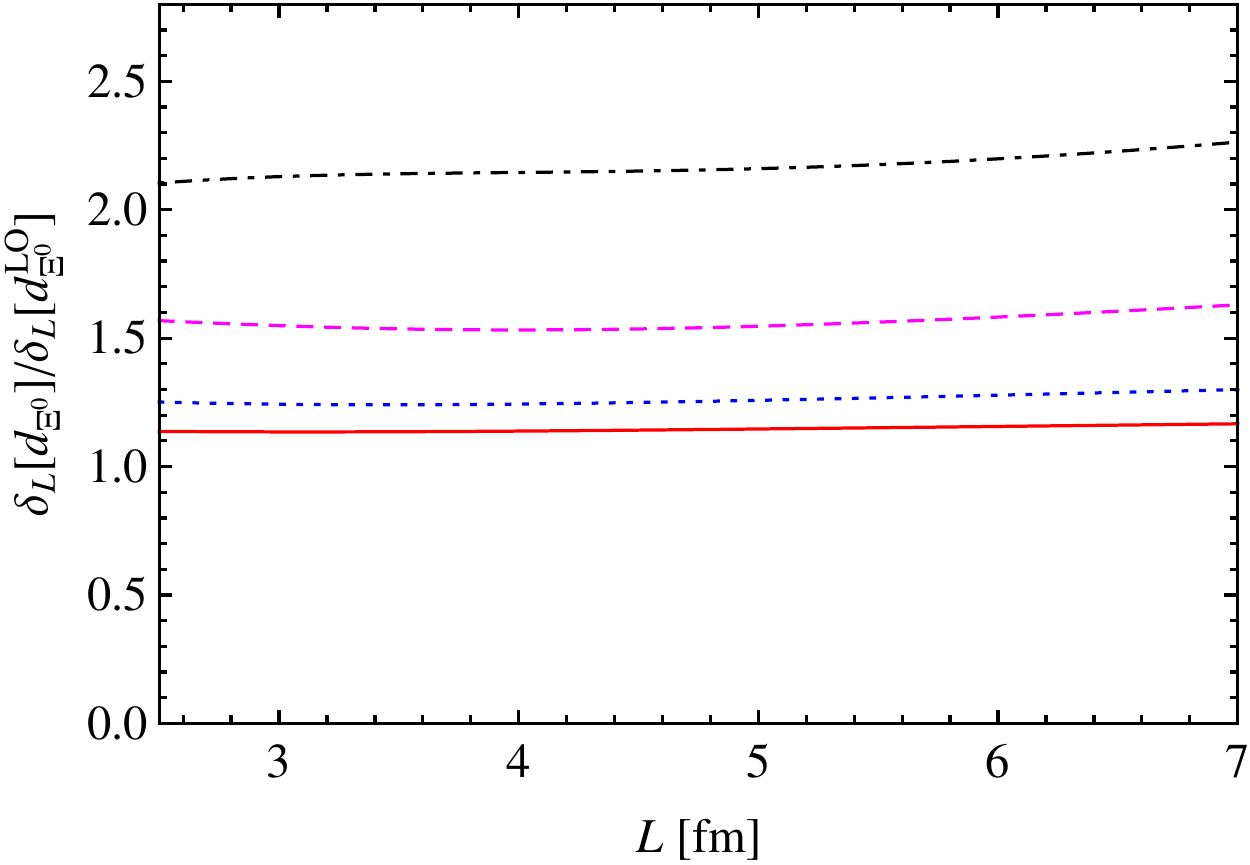}
\caption{Left: The ratios of the finite volume corrections to the loop contributions
in the infinite volume for the EDMs of neutral hyperons as a function of $L$. Right: The
ratios of the finite volume corrections of NLO to those of LO. The solid, dotted,
dashed and dot-dashed lines are for the pion mass being 138~Nev (physical
value), 200~MeV, 300~MeV and 400~MeV, respectively.
\label{fig:HEDMFV}}
\end{figure}
The left panel of figure~\ref{fig:nEDMFV} shows the
ratios of the finite volume corrections to the loop contributions in the
infinite volume  for the neutron EDM as a function of $L$. The results are
obtained with $\mu=1$~GeV. In order to show the impact of the
NLO corrections more clearly, we plot the ratios of the finite volume
corrections at  NLO (LO not included) to the LO in the right panel of
figure~\ref{fig:nEDMFV}.
One sees that at the physical pion mass the NLO contributions
increase the LO finite volume corrections by
about 20\%. This correction is  much larger for a pion mass of 400~MeV.
At this point, one might worry about the convergence. In fact, the
neutron EDM in the infinite volume
has a nice convergence property, as can be seen from the left panel of
figure~\ref{fig:dn}. Moreover, one can show that the ratio tends to increase with
increasing $L$. Although both the LO and NLO finite volume corrections decrease
exponentially, the LO one decreases faster. Using eqs.~\eqref{eq:DeltaLJm1m2Asymp}
and \eqref{eq:DeltaLJMMmAsymp} from appendix~\ref{app:FVasy}, we get in
the limit $L \to \infty$,
\begin{equation}
   \frac{\delta_L \left[ d_n^\text{NLO} \right]} {\delta_L \left[ d_n^\text{LO}
   \right]} \sim \sqrt{\frac\pi2} \frac{M_\pi}{m_N} \sqrt{LM_\pi} \exp\left(
   \frac{LM_\pi^3}{8 m_N^2}
   \right).
\end{equation}
The finite volume corrections to the neutral hyperons are shown in
figure~\ref{fig:HEDMFV}. It is obvious that the effects are very small for the
$\Lambda$ and $\Sigma^0$. The reason is that the pionic loops with positively and
negatively charged pions cancel each other exactly for these two baryons. This also
makes the NLO corrections much larger than the LO ones at $\mu=1$~GeV, consistent
with the infinite volume results shown in figure~\ref{fig:dn}.
The situation for the $\Xi^0$ is similar to case of the neutron.

\begin{figure}[t]
\centering
\includegraphics[width=\linewidth]{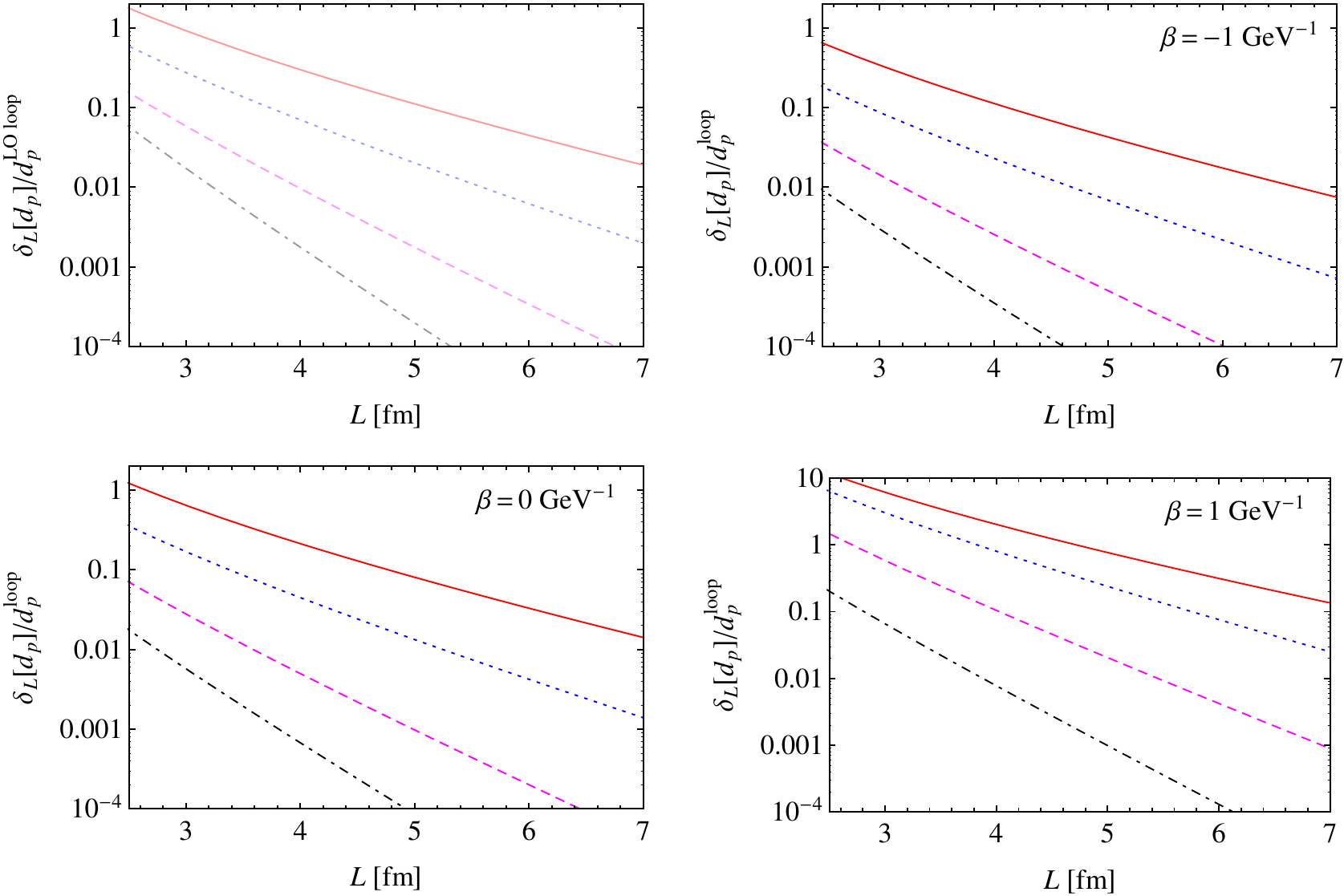}
\caption{The ratio of the finite volume corrections to the loop contributions
in the infinite volume for the proton EDM as a function of $L$. The first graph presents
the LO results, and the others are the NLO results for different choices of $\beta$. The
solid, dotted, dashed and dot-dashed lines are for the pion mass of 138~MeV,
200~MeV, 300~MeV and 400~MeV, respectively.
\label{fig:pEDMFV}}
\end{figure}

Although the parameter combination $w_b(\mu)$ can be determined from the lattice
data of the proton EDM, $\beta$ is still unknown since it always appear together
with the LECs $w_{14}$ and $w_{14}^{\prime\,r}(\mu)$. However, the finite volume
corrections for the charged baryons depend on $\beta$, but not on $w_{14}$ and
$w_{14}^{\prime\,r}(\mu)$. It is therefore possible to extract the value of $\beta$
from the volume effects of the baryon EDMs. In figure~\ref{fig:pEDMFV}, we show the
ratios of the finite volume corrections to the loop contributions in infinite
volume to the proton EDM for different chosen values of $\beta$. At this point, we
want to emphasize that the EDMs of all the charged baryons depend on the same
combination of LECs, i.e. $\beta$. Once it is determined from one baryon, it can be
used for predicting the finite volume corrections to the others.

\section{Summary}
\label{sec:sum}

In this paper, we extended the previous calculation~\cite{Ottnad:2009jw} of the
nucleon EDFFs and EDMs to the ground state baryon octet up to NLO in the
framework of U(3)$_L \times$U(3)$_R$ chiral perturbation theory. Our main
findings can be summarized as follows:
\begin{itemize}
\item[1)] We have shown that the complete one-loop expressions for the baryon
  EDFFs and EDMs depend on two combinations of unknown LECs only, $w_a$ and
  $w_b$, cf. eqs.(\ref{eq:w13},\ref{eq:w14beta}). In case of the charged baryons,
  the combination $w_b(\mu)$ combines two LECs from the tree graphs and
  one LEC that appears only in loops.
\item[2)] We have shown that the NLO corrections are large for the neutral
    hyperons $\Lambda$ and $\Sigma^0$. This is due to a suppression of the LO
    contributions based on exact cancellations between loops of positively and
    negatively charged pions. For the charged baryons, we find a strong
    sensitivity to the LEC combination $w_b$.
\item[3)] We have derived a set of relations between various EDMs that are
  free of unknown LECs. These can be useful for future lattice simulations
  of baryon EDMs.
\item[4)] Based on recent lattice results for the neutron and proton EDMs at
    $M_\pi = 530\,$MeV, we could pin down the two LEC combinations $w_a$ and
    $w_b$. Based on this, we can predict the baryon EDMs at the physical pion
    mass. In particular, we find $d_n = -2.9\pm 0.9$ and $d_p = 1.1\pm 1.1$ in
    units of $10^{-16}\,\theta_0\, e$~cm.
\item[5)]The finite volume corrections to the baryon EDMs in the $p$-regime are
    also studied. Because the loops contribute from LO, it is found that the
    finite volume corrections are huge for all the baryon EDMs except for the
    $\Lambda$ and $\Sigma^0$ for the pion mass close to its physical value. For
    the neutron, the finite volume correction is about 10\% at $M_\pi L=4$.
\end{itemize}

Note that the calculation of the finite volume corrections in our paper assumes
Lorentz invariance, and additional subtleties in the presence of external
electromagnetic fields, see~\cite{Tiburzi:2008pa}, are not taken into account. For
a precise calculation of the finite volume corrections of the baryon EDMs, separate
analyses need to be done for the lattice calculations using the form factor method,
the spectrum method with external field and methods with the twisted boundary
conditions.

More lattice results of the nucleon EDMs are expected to come out soon.
Calculations of the hyperons are welcome based on our analysis. With these upcoming
lattice data, one can determine the LECs more reliably, and a
next-to-next-to-leading order calculation would be feasible and desirable in view
of the bad convergence for some baryons. Furthermore, it would be interesting to
see how the LEC-free relations get modified at higher orders.

\medskip

\acknowledgments{
 We are grateful to E.~Shintani for providing us with lattice data
before publication. We acknowledge discussions with T.~Izubichi and G.~Schierholz.
We also thank B. C. Tiburzi for very useful comments on the finite volume
corrections. This work is supported in part by the DFG and the NSFC through funds
provided to the Sino-German CRC 110 ``Symmetries and the Emergence of Structure in
QCD'' and the EU I3HP ``Study of Strongly Interacting Matter'' under the Seventh
Framework Program of the EU. U.-G. M. also thanks the BMBF for support (Grant No.
06BN7008). F.-K. G. acknowledges partial support from the NSFC (Grant No.
11165005). }

\medskip

\appendix

\section{Baryon masses up to NLO}
\label{app:mb}

From eq.~\eqref{eq:BaryonLagrangian}, it is easy to obtain the baryon masses up to
$\order{2}$, which may be written as
\begin{equation}
   m_B = \mathring{m} - 2 b_0 \left( M_\pi^2 + 2 M_K^2 \right) - 4 \left( b_\pi
   M_\pi^2 + b_K M_K^2 \right) + \order{3},
   \label{eq:mb}
\end{equation}
where we have expressed the quark masses in terms of the meson masses at the lowest
order, and $b_{\pi,K}$ are listed in table~\ref{tab:mb}.
\begin{table}[t]
\begin{center}
   \begin{tabular}{|l | c c c c |} \hline
      Baryons & $N$       & $\Sigma$ & $\Lambda$ & $\Xi$     \\
      \hline
      $b_\pi$ & $b_F$     & $b_D$    & $-b_D/3$  & $-b_F$    \\
      $b_K$   & $b_D-b_F$ & 0        & $4b_D/3$ & $b_D+b_F$ \\
      \hline
   \end{tabular}
   \caption{\label{tab:mb} Coefficients $b_{\pi,K}$ in the NLO baryon mass formula.}
\end{center}
\end{table}

\section{Loop integrals in infrared regularization}
\label{app:loop}

Loop integrals in the infinite volume used in this paper are collected in this
appendix. Let us denote the masses of the light meson and baryon in the loop by $M$
and $m$, respectively. The momentum of the external current is $q^\mu$, and the
momenta of the external baryons are denoted by $p^\mu$ and $p^{\prime
\mu}=p^\mu+q^\mu$, respectively. We further define
$$L = \frac{\mu^{d-4}}{(4\pi)^2} \left\{ \frac1{d-4} -\frac12\left[ \ln (4\pi)
   + \varGamma'(1) + 1 \right] \right\},$$
which contains the divergence at the spacetime dimension $d=4$.

The 2-point mesonic-loop function when the two mesons have the same mass reads as
\begin{eqnarray}
      J_{MM}(q^2) \al=\al i\int \frac{d^dk}{(2\pi)^d} \frac1{(k^2-M^2+i\,\epsilon)
      \left[(k+q)^2-M^2+i\epsilon\right]} \nonumber\\
      \al=\al 2 L + \frac1{16\pi^2}\left( \ln\frac{M^2}{\mu^2} -1 - \sigma
      \ln\frac{\sigma-1}{\sigma+1} \right),
      \label{eq:JMM}
\end{eqnarray}
where $\sigma=\sqrt{1-4M^2/q^2}$.

When there is a baryon in the loop, the loop evaluated in the normal dimensional
regularization spoils power counting~\cite{Gasser:1987rb}. The power counting can
be restored in the heavy baryon formalism of baryon
CHPT~\cite{Jenkins:1990jv,Bernard:1995dp}. There are also covariant
formalisms, see ref.~\cite{Bernard:2007zu} for a recent review.

Here, the infrared regularization~\cite{Ellis:1997kc,Becher:1999he} will be used.
In the infrared regularization, the integral is separated into an infrared regular
piece and an infrared singular piece. The infrared regular piece can be expanded
analytically in the chiral expansion, and absorbed into the counterterms. Only the
infrared singular part of the loop integrals will be shown in the following.

The two-point scalar loop integral involving one meson and one baryon is
\begin{eqnarray}
      J_{Mm}(p^2) \al=\al i\int \frac{d^dk}{(2\pi)^d} \frac1{(k^2-M^2+i\,\epsilon)
      \left[(p-k)^2-m^2+i\epsilon\right]} \nonumber\\
      \al=\al \frac{\omega}{p^2} L + \frac1{32\pi^2p^2} \left[
      \omega\left(\ln \frac{M^2}{\mu^2}-1\right) + 2
      \sqrt{\lambda(p^2,m^2,M^2)} \arccos\left(-\frac{\omega}{2\tilde pm}\right)
      \right],
\end{eqnarray}
where $\omega=p^2-m^2+M^2$, $\tilde p=\sqrt{p^2}$, and
$\lambda(x,y,z)=x^2+y^2+z^2-2(xy+yz+xz)$ is the K\"all\'en function. Noting that
the external baryons are on shell, $p^2$ is given by the mass squared of the
external baryon. Thus, with the baryon mass expressions in appendix~\ref{app:mb},
one has $p^2=m^2+\order{2}$ and $\omega=\order{2}$. Expanding the $\arccos$
function, we get
\begin{eqnarray}
      J_{Mm}(p^2) \al=\al \frac{M}{16\pi\, \tilde p} + \frac{\omega}{p^2}\left[ L +
      \frac1{32\pi^2} \left( 1 + \ln \frac{M^2}{\mu^2} \right) \right] + \order{3},
      \label{eq:JMm2}
\end{eqnarray}
where the first and second term are of $\mathcal{O}(\delta)$ and $\order{2}$,
respectively.

The three-point loop integral with two mesons and one baryon can be written as
\begin{eqnarray}
      J_{MMm}(q^2,p^2) \al=\al i\int \frac{d^dk}{(2\pi)^d}
      \frac1{(k^2-M^2+i\,\epsilon) \left[(k+q)^2-M^2+i\epsilon\right]
      \left[(p-k)^2-m^2+i\,\epsilon\right]} \non\\
      \al = \al \frac{\partial}{\partial M^2} \int_0^1\!\!dx J_{\bar M m}(\bar p^2),
      \label{eq:JMMm}
\end{eqnarray}
where $\bar M^2 = M^2+x(x-1)q^2$, $\bar p^\mu = p^\mu + x\,q^\mu$. In the
following, $p^{\prime\, 2}=p^2$ will always be assumed, then one has $\bar p^2 =
p^2 + x(x-1)q^2$. The leading chiral order of this integral is
$\mathcal{O}(\delta^{-1})$. Keeping terms up to $\mathcal{O}(\delta^0)$, the
analytic expression for the infrared singular part is
\begin{eqnarray}
      J_{MMm}(q^2,p^2) \al=\al \frac1{16\pi\, \tilde p\sqrt{-q^2}} \arctan
      \frac{\sqrt{-q^2}}{2M} \nonumber \\
      \al\al + \left[ \frac{L}{p^2} + \frac1{32\pi^2p^2} \left( 1 + \ln
      \frac{M^2}{\mu^2} + \frac{2\omega-q^2}{q^2\sigma} \ln
      \frac{\sigma-1}{\sigma+1} \right) \right] + \mathcal{O}(\delta),
      \label{eq:JMMm2}
\end{eqnarray}
where the first term is of order $\mathcal{O}(\delta^{-1})$, and the terms in the
square brackets are of order $\mathcal{O}(\delta^0)$.

For completeness, we also give the expression for the loop with one meson and two
baryons, though not used in our calculations. Similar to $J_{MMm}(q^2,p^2)$, this
loop can also be worked out from the two-point loop with one meson and one baryon,
\begin{eqnarray}
      J_{Mmm}(p^2,p^{\prime 2}) \al=\al i\int \frac{d^dk}{(2\pi)^d}
      \frac1{(k^2-M^2+i\,\epsilon) \left[(p-k)^2-m^2+i\,\epsilon\right]
      \left[(p'-k)^2-m^2+i\,\epsilon\right]} \nonumber\\
      \al = \al \frac{\partial}{\partial m^2} \int_0^1\!\!dx J_{M\bar m}(\bar p^2),
      \label{eq:JMmm}
\end{eqnarray}
where $\bar m^2= m^2+x(x-1)q^2$. The analytic expression up to NLO is
\begin{eqnarray}
      J_{Mmm}(p^2,p^{\prime 2}) =  - \left[ \frac{L}{p^2} + \frac1{32\pi^2p^2}
      \left( 1 + \ln \frac{M^2}{\mu^2} \right) \right] + \frac{\omega}{64\pi p^2 mM} + \order{2},
      \label{eq:JMmm2}
\end{eqnarray}
where the terms in the square brackets are at LO, and the second term is at NLO. Up
to NLO, this loop does not depend on the momentum transfer $q^2$.

\section{Expressions for the baryon EDFFs up to NLO}
\label{app:expressions}

Up to NLO, which is $\order{3}$, the baryon EDFFs contain two parts: the tree-level
expressions are given in table~\ref{tab:tree}, and the loop contributions are
obtained by taking the LO terms of the baryonic loops $J_{MM\tilde m}(q^2,m^2)$ and
$J_{M\tilde m}(q^2)$. The explicit expressions of the EDFFs of the baryon octet up
to NLO are given in the following:
\begin{itemize}[itemindent=4mm,leftmargin=0mm]
\item For the neutron,
      \begin{eqnarray}
         \frac{F_{3,n}(q^2)}{2 m_N} \al=\al \frac83 e \bar{\theta }_0 \left[\alpha
         w_{13}+w_{13}^{\prime\, r}(\mu)\right] %
         + \frac{V_0^{(2)} e \bar\theta_0}{\pi^2 F_\pi^4} \Bigg\{
         (D+F)\left(b_D+b_F\right) \Bigg[ 1 - \ln \frac{M_\pi^2}{\mu^2}
         + \sigma_\pi \ln\frac{\sigma_\pi-1}{\sigma_\pi+1}  \nonumber\\
         \al\al +
         \frac{\pi \left(2M_\pi^2-q^2\right)} {2m_N\sqrt{-q^2}} \arctan
         \frac{\sqrt{-q^2}}{2M_\pi} \Bigg] \! - (D-F)\left(b_D-b_F\right) \Bigg[ 1
         - \ln \frac{M_K^2}{\mu^2} + \sigma_K \ln \frac{\sigma_K-1}{\sigma_K+1}
          \nonumber\\
         \al\al + \frac{\pi}{\sqrt{-q^2}} \left( \frac{2 M_K^2-q^2}{2
         m_N} - 8 \left(b_D-b_F\right) \left(M_K^2-M_\pi^2\right) \right) \arctan
         \frac{\sqrt{-q^2}}{2 M_K} \Bigg] \Bigg\},
         \label{eq:F3n}
      \end{eqnarray}
      with $\alpha = 144 V_0^{(2)} V_3^{(1)}/(F_0 F_\pi M_{\eta_0})^2$, and
      $\sigma_{\pi(K)}=\sqrt{1-4M_{\pi(K)}^2/q^2}$. Here, we have used the
      SU(3) mass splitting for the baryon masses $m_\Sigma-m_N=4
      \left(b_D-b_F\right) \left(M_K^2-M_\pi^2\right) + \order{3}$. Similar
      relations derived from the Lagrangian eq.~\eqref{eq:BaryonLagrangian}
      will be used in the following.
\item For the proton,
      \begin{eqnarray}
         \frac{F_{3,p}(q^2)}{2 m_N} \al=\al -\frac43 e \bar{\theta }_0 \left[\alpha
         \left(w_{13}+3 w_{14}\right)+w_{13}^{\prime\, r}(\mu)+3 w_{14}^{\prime\,
         r}(\mu) \right] - \frac{V_0^{(2)} e \bar\theta_0}{6\pi^2 F_\pi^4} \Bigg\{ 6
         (D+F) \left(b_D+b_F\right) \nonumber\\
         \al\al \times \Bigg[ 1 - \ln \frac{M_\pi^2}{\mu^2} + \sigma_\pi \ln
         \frac{\sigma_\pi-1}{\sigma_\pi+1} + \frac{3\pi
         M_\pi}{2 m_N} + \frac{\pi \left(2M_\pi^2-q^2\right)}{2m_N\sqrt{-q^2}}
         \arctan\frac{\sqrt{-q^2}}{2M_\pi} \Bigg] \nonumber\\
         \al\al + 4 \left(Db_D + 3Fb_F\right) \bigg( 1 - \ln \frac{M_K^2}{\mu^2} +
         \sigma_K \ln \frac{\sigma_K-1}{\sigma_K+1} +
         \frac{\pi M_K}{m_N} \bigg) \nonumber\\
         \al\al  + \frac{4\pi}{\sqrt{-q^2}}
         \arctan\frac{\sqrt{-q^2}}{2 M_K} \bigg[ \frac{\left(Db_D +
         3Fb_F\right)}{2m_N} \left(2 M_K^2-q^2\right) \nonumber\\
         \al\al
         + 8 \left(M_K^2-M_\pi^2\right) \left( F \left(b_D^2+3 b_F^2\right)-\frac23 D
         b_D \left(b_D-3 b_F\right) \right)
         \bigg] \nonumber\\
         \al\al + \frac{\pi}{m_N} \bigg[ 6 (D-F) \left(b_D-b_F\right) M_K + (D-3F)
         \left(b_D-3b_F\right) M_{\eta_8} + \frac{2F_\pi^2}{F_0^2} \beta M_{\eta_0} \bigg]
          \Bigg\}, \nonumber \\
          \label{eq:EDFFp}
      \end{eqnarray}
      where $\beta = (2D-3w_0) \left(2b_D+3b_0+6w_{10}'\right)$.
\item For the $\Sigma^0$,
      \begin{eqnarray}
         \frac{F_{3,\Sigma^0}(q^2)}{2 m_\Sigma} \al=\al -\frac43 e \bar{\theta
         }_0 \left[\alpha w_{13}+w_{13}^{\prime\, r}(\mu) \right] %
         - \frac{V_0^{(2)} e \bar\theta_0}{\pi^2 F_\pi^4} \Bigg\{
         \left(Db_F+Fb_D\right) \bigg( 1 - \ln\frac{M_K^2}{\mu^2} \nonumber\\
         \al\al + \sigma_K \ln\frac{\sigma_K-1}{\sigma_K+1} \bigg) +
         \frac{\pi}{\sqrt{-q^2}} \arctan\frac{\sqrt{-q^2}}{2 M_K} \bigg[
         \frac{Db_F+Fb_D}{2 m_\Sigma} \left(2M_K^2-q^2\right) \nonumber\\
         \al\al + 8 \left(M_K^2-M_\pi^2\right) \left(F b_D^2+2 D b_D b_F+Fb_F^2\right)
         \bigg] \Bigg\}.
      \end{eqnarray}
\item For the $\Sigma^+$,
      \begin{eqnarray}
         \frac{F_{3,\Sigma^+}(q^2)}{2 m_\Sigma} \al=\al -\frac43 e \bar{\theta }_0
         \left[\alpha \left(w_{13}+3 w_{14}\right)+w_{13}^{\prime\, r}(\mu)+3
         w_{14}^{\prime\, r}(\mu) \right]  \nonumber\\
         \al\al - \frac{V_0^{(2)} e \bar\theta_0}{3\pi^2
         F_\pi^4} \Bigg\{ 2\left( Db_D+3Fb_F \right) \bigg( 1 - \ln \frac{M_\pi^2}{\mu^2}
         + \sigma_\pi \ln
         \frac{\sigma_\pi-1}{\sigma_\pi+1} + \frac{\pi
         M_\pi}{m_\Sigma} \bigg) \nonumber\\
         \al\al + 3 (D+F) \left(b_D+b_F\right) \Bigg[ 1
         - \ln\frac{M_K^2}{\mu^2} + \sigma_K \ln \frac{\sigma_K-1}{\sigma_K+1} +
         \frac{\pi M_K}{m_\Sigma} \nonumber\\
         \al\al + \frac{\pi}{\sqrt{-q^2}} \left( \frac{2 M_K^2-q^2}{2
         m_\Sigma} + 8 \left(b_D+b_F\right) \left(M_K^2-M_\pi^2\right) \right)
         \arctan\frac{\sqrt{-q^2}}{2 M_K} \Bigg] \nonumber\\
         \al\al + \frac{2\pi} {\sqrt{-q^2}}
         \arctan\frac{\sqrt{-q^2}}{2M_\pi} \bigg[
         \frac{\left(Db_D+3Fb_F\right)}{2m_\Sigma} \left(2M_\pi^2-q^2\right) +
         \frac{32}{3} D b_D^2 \left(M_K^2-M_\pi^2\right) \bigg]
         \nonumber\\
         \al\al + \frac{\pi}{m_\Sigma} \bigg[ 6Fb_F M_\pi + 3 (D-F)
         \left(b_D-b_F\right) M_K + 2Db_D M_{\eta_8} + \frac{F_\pi^2}{F_0^2} \beta
         M_{\eta_0} \bigg] \Bigg\}.
         \label{eq:EDFFSp}
      \end{eqnarray}
\item For the $\Sigma^-$,
      \begin{eqnarray}
         \frac{F_{3,\Sigma^-}(q^2)}{2 m_\Sigma} \al=\al -\frac43 e \bar{\theta }_0
         \left[\alpha \left(w_{13}-3 w_{14}\right)+w_{13}^{\prime\, r}(\mu)-3
         w_{14}^{\prime\, r}(\mu) \right] \nonumber\\
         \al\al + \frac{V_0^{(2)} e \bar\theta_0}{3\pi^2 F_\pi^4} \Bigg\{ 2\left(
         Db_D+3Fb_F \right) \left( 1 - \ln \frac{M_\pi^2}{\mu^2} + \sigma_\pi \ln
         \frac{\sigma_\pi-1}{\sigma_\pi+1} + \frac{\pi
         M_\pi}{m_\Sigma} \right) \nonumber\\
         \al\al + \frac{2\pi} {\sqrt{-q^2}}
         \arctan\frac{\sqrt{-q^2}}{2M_\pi} \bigg[
         \frac{\left(Db_D+3Fb_F\right)}{2m_\Sigma} \left(2M_\pi^2-q^2\right) +
         \frac{32}{3} D b_D^2 \left(M_K^2-M_\pi^2\right) \bigg]  \nonumber\\
         \al\al + 3 (D-F) \left(b_D-b_F\right) \Bigg[ 1 - \ln \frac{M_K^2}{\mu^2} +
         \sigma_K \ln \frac{\sigma_K-1}{\sigma_K+1} + \frac{\pi M_K}{m_\Sigma} \nonumber\\
         \al\al +
         \frac{\pi}{\sqrt{-q^2}} \left( \frac{2 M_K^2-q^2}{2
         m_\Sigma} + 8 \left(b_D-b_F\right) \left(M_K^2-M_\pi^2\right) \right)
         \arctan\frac{\sqrt{-q^2}}{2 M_K} \Bigg] \nonumber\\
         \al\al + \frac{\pi}{m_\Sigma} \bigg[ 6Fb_F M_\pi + 3 (D+F)
         \left(b_D+b_F\right) M_K + 2Db_D M_{\eta_8} + \frac{F_\pi^2}{F_0^2} \beta
         M_{\eta_0} \bigg] \Bigg\}.
         \label{eq:EDFFSm}
      \end{eqnarray}
\item For the $\Lambda$
      \begin{eqnarray}
         \frac{F_{3,\Lambda}(q^2)}{2 m_\Lambda} \al=\al \frac43 e \bar{\theta }_0
         \left[\alpha w_{13}+w_{13}^{\prime\, r}(\mu) \right] %
         + \frac{V_0^{(2)} e \bar\theta_0}{\pi^2 F_\pi^4} \Bigg\{
         \left(Db_F+Fb_D\right) \left( 1 - \ln\frac{M_K^2}{\mu^2} + \sigma_K \ln
         \frac{\sigma_K-1}{\sigma_K+1} \right) \nonumber\\
         \al\al + \frac{\pi}{\sqrt{-q^2}} \arctan\frac{\sqrt{-q^2}}{2 M_K}
         \bigg[ \frac{Db_F+Fb_D}{2 m_\Sigma} \left(2M_K^2-q^2\right) \nonumber\\
         \al\al - \frac83 \left(M_K^2-M_\pi^2\right) \left(F b_D^2+2 D b_D
         b_F+Fb_F^2\right) \bigg] \Bigg\}.
      \end{eqnarray}
\item For the $\Xi^0$,
      \begin{eqnarray}
         \frac{F_{3,\Xi^0}(q^2)}{2 m_\Xi} \al=\al \frac83 e \bar{\theta }_0
         \left[\alpha w_{13}+w_{13}^{\prime\, r}(\mu) \right] %
         - \frac{V_0^{(2)} e \bar\theta_0}{\pi^2 F_\pi^4} \Bigg\{
         (D-F)\left(b_D-b_F\right) \Bigg[ 1 - \ln \frac{M_\pi^2}{\mu^2} + \sigma_\pi
         \ln\frac{\sigma_\pi-1}{\sigma_\pi+1}  \nonumber\\
         \al\al +
         \frac{\pi \left(2M_\pi^2-q^2\right)} {2m_\Xi\sqrt{-q^2}} \arctan
         \frac{\sqrt{-q^2}}{2M_\pi} \Bigg] \! - (D+F)\left(b_D+b_F\right)
         \Bigg[ 1 - \ln \frac{M_K^2}{\mu^2} + \sigma_K \ln
         \frac{\sigma_K-1}{\sigma_K+1} \nonumber\\
         \al\al  + \frac{\pi}{\sqrt{-q^2}} \left( \frac{2 M_K^2-q^2}{2
         m_\Xi} - 8 \left(b_D+b_F\right) \left(M_K^2-M_\pi^2\right) \right) \arctan
         \frac{\sqrt{-q^2}}{2 M_K} \Bigg] \Bigg\},
      \end{eqnarray}
\item For the $\Xi^-$,
      \begin{eqnarray}
         \frac{F_{3,\Xi^-}(q^2)}{2 m_\Xi} \al=\al -\frac43 e \bar{\theta }_0
         \left[\alpha \left(w_{13}-3 w_{14}\right)+w_{13}^{\prime\, r}(\mu)-3
         w_{14}^{\prime\, r}(\mu) \right] %
         + \frac{V_0^{(2)} e \bar\theta_0}{6\pi^2 F_\pi^4} \Bigg\{ 6
         (D-F) \left(b_D-b_F\right) \nonumber\\
         \al\al \times \Bigg[ 1 - \ln \frac{M_\pi^2}{\mu^2} + \sigma_\pi \ln
         \frac{\sigma_\pi-1}{\sigma_\pi+1} + \frac{3\pi
         M_\pi}{2 m_\Xi} + \frac{\pi \left(2M_\pi^2-q^2\right)}{2m_\Xi\sqrt{-q^2}}
         \arctan\frac{\sqrt{-q^2}}{2M_\pi} \Bigg] \nonumber\\
         \al\al + 4 \left(Db_D + 3Fb_F\right) \bigg( 1 - \ln \frac{M_K^2}{\mu^2} +
         \sigma_K \ln \frac{\sigma_K-1}{\sigma_K+1} +
         \frac{\pi M_K}{m_\Xi} \bigg) \nonumber\\
         \al\al  + \frac{4\pi}{\sqrt{-q^2}}
         \arctan\frac{\sqrt{-q^2}}{2 M_K} \bigg[ \frac{Db_D+3Fb_F}{2m_\Xi}
         \left(2 M_K^2-q^2\right) \nonumber\\
         \al\al
         - 8 \left(M_K^2-M_\pi^2\right) \left( F \left(b_D^2+3 b_F^2\right)+\frac23 D
         b_D \left(b_D+3 b_F\right) \right) \bigg] \nonumber\\
         \al\al + \frac{\pi}{m_\Xi} \bigg[ 6 (D+F) \left(b_D+b_F\right) M_K + (D+3F)
         \left(b_D+3b_F\right) M_{\eta_8} + \frac{2F_\pi^2}{F_0^2} \beta M_{\eta_0} \bigg]
          \Bigg\}. \nonumber\\
          \label{eq:EDFFXm}
      \end{eqnarray}
\end{itemize}

The expressions for the EDMs can be easily obtained by noticing
$$ \lim_{q^2\to 0} \sigma \ln \frac{\sigma-1}{\sigma+1} = -2, \qquad
\lim_{q^2\to 0} \frac1{\sqrt{-q^2}} \arctan\frac{\sqrt{-q^2}}{2 M} = \frac1{2M} .
$$

\section{Finite volume corrections to loops}
\label{app:FVloop}

This appendix is dedicated to finite volume corrections to the two- and three-point
loop integrals. Let us consider the scalar two-point loop
\begin{eqnarray}
      J_{m_1m_2}(q^2) \al=\al i\int \frac{d^4k}{(2\pi)^4}
      \frac1{(k^2-m_1^2+i\,\epsilon) \left[(k+q)^2-m_2^2+i\epsilon\right]} \non\\
      \al=\al \int_0^1\!\! dx\, i \int \frac{d^4k}{(2\pi)^4} \frac1{\left( k^2 -
      \bar m_{12}^2 + i\epsilon \right)^2} ,
\end{eqnarray}
where $\bar m_{12}^2 = x(x-1)q^2+(1-x)m_1^2+x\,m_2^2$. If the loop involves both a
meson and a baryon, $m_1$ should be replaced by the mesonic mass, and the upper
bound of the integration over the Feynman parameter $x$ should be replaced by
$\infty$ in infrared regularization. Performing the contour integral over $k^0$,
one gets
\begin{equation}
   J_{m_1m_2}(q^2) = - \frac14 \int_0^1\!\! dx \int \frac{d^3\vec{k}}{(2\pi)^3}
   \frac1{\left( \vec{k}^{\,2} + \bar m_{12}^2 \right)^{3/2}}.
\end{equation}
Using the formula
\begin{eqnarray}
   \left(\frac1{L^3} \sum_{\vec{n}} - \int \frac{d^3\vec{k}}{(2\pi)^3}
   \right)\frac1{\left(\vec{k}^2+a^2\right)^j} =
   \frac{2^{-j}a^{3-2j}}{\sqrt{2}\pi^{3/2}\Gamma(j)} \sum_{\vec{n}\neq0}
   \frac{K_{3/2-j}(L\,a|\vec{n}|)}{(L\,a|\vec{n}|)^{3/2-j}}
\end{eqnarray}
derived in ref.~\cite{Beane:2004tw}, we get the finite volume correction to the
loop $J_{m_1m_2}(q^2)$,
\begin{equation}
   \delta_L [J_{m_1m_2}(q^2)] = - \frac1{8\pi^2} \int_0^1\!\!dx \sum_{\vec{n}\neq0}
   K_0(L \bar m_{12} |\vec{n}|),
   \label{eq:DeltaLJm1m2}
\end{equation}
with $K_\nu(z)$ a modified Bessel function of the second kind.

From eq.~\eqref{eq:JMMm}, it is easy to write down the correction to the
three-point loop integral,
\begin{eqnarray}
   \delta_L [J_{MMm}(q^2,p^2)] \al=\al \frac{\partial}{\partial M^2} \int_0^1\!\!dx
   \, \delta_L [J_{\bar M m}(\bar p^2)] \nonumber\\
   \al=\al \frac{L^2}{16\pi^2} \sum_{\vec{n}\neq0} \vec{n}^{\,2} \int_0^1\!\!dx
   \int_0^\infty dy \frac{1-y}{z_1} K_1(z_1),
   \label{eq:DeltaLJMMm}
\end{eqnarray}
where $z_1=L|\vec{n}|\left[y(y-1)\bar p^2+y m^2+(1-y)\bar M^2\right]^{1/2}$. We
have made use of the integral representation of $K_n(z)$~\cite{handbook}
\begin{equation}
   K_n(z) = \frac{\Gamma(n+1/2) (2z)^n}{\sqrt{\pi}} \int_0^\infty \!\! dt \frac{\cos
   t}{\left(t^2+z^2\right)^{n+1/2}}.
\end{equation}
which is valid for ${\rm Re}\, n>-1/2, |\arg z|<\pi/2$. Similarly, one gets
\begin{eqnarray}
   \delta_L [J_{Mmm}(p^2,p^{\prime2})] = \frac{L^2}{16\pi^2} \sum_{\vec{n}\neq0}
   \vec{n}^{\,2} \int_0^1\!\!dx \int_0^\infty dy \frac{y}{z_2} K_1(z_2),
   \label{eq:DeltaLJMmm}
\end{eqnarray}
where $z_2=L|\vec{n}|\left[y(y-1)\bar p^2+y \bar m^2+(1-y)M^2\right]^{1/2}$.

\section{Asymptotic expansion of finite volume corrections}
\label{app:FVasy}

For simplicity, we will focus on the asymptotic expansion of finite volume
corrections with $q^2=0$ and $p^2=m^2$. The general case can be treated similarly.
For $|z|\to \infty$, one has
\begin{equation}
   K_n(z) \sim \sqrt{\frac\pi{2z}} e^{-z} \left[ 1 + \mathcal{O}\left( \frac1z
   \right) \right].
\end{equation}
For $L\to \infty$, we can consider the term with $|\vec{n}|=1$ only in
eq.~\eqref{eq:DeltaLJm1m2}. There are 6 possibilities, so that for $L\to \infty$,
\begin{equation}
   \delta_L [J_{MM}(0)] \sim - \frac3{4\pi^2} \sqrt{\frac\pi{2}}
   \frac{ e^{-L M} }{\sqrt{L M}}.
   \label{eq:DeltaLJm1m2Asymp}
\end{equation}
For the three-point loop integral,
\begin{eqnarray}
   \delta_L [J_{MMm}(0,m^2)] \sim \frac{L^2}{16\pi^2} \sqrt{\frac\pi{2}}
   \int_0^\infty dy\, (1-y) \frac{ e^{-L C(y)} }{\left[L C(y)\right]^{3/2}},
\end{eqnarray}
with $C(y)=[y^2 m^2+(1-y)M^2]^{1/2}$. The leading term of the above integral over
the Feynman parameter $y$ can be worked out using Laplace's method. For very large
$L$, the integral receives contributions mostly from the neighborhood of the
minimum of $C(y)$, which is at $y_c = M^2/(2m^2)$. Thus,
\begin{eqnarray}
   \delta_L [J_{MMm}(0,m^2)] \al\sim\al \frac{3L^2}{8\pi^2} \sqrt{\frac\pi{2}}
   (1-y_c) \frac{ e^{-L C(y_c)} }{\left[L C(y_c)\right]^{3/2}}
   \int_{-\infty}^\infty dy \, \exp\left[-\frac12 y^2 C''(y_c)\right] \nonumber \\
   \al = \al \frac3{8\pi mM} \exp\left( - LM\sqrt{1-\frac{M^2}{4 m^2}}
   \right) \left[1 + \mathcal{O} \left(\frac{M^2}{m^2}\right) \right].
   \label{eq:DeltaLJMMmAsymp}
\end{eqnarray}
The asymptotic expansion for $\delta_L [J_{Mmm}(m^2,m^2)]$ is the same.


\begin{thebibliography}{99}

%
\bibitem{Baker:2006ts}
  C.~A.~Baker et al.,
  {\it An improved experimental limit on the electric dipole moment of the
  neutron},
  {\it Phys.\ Rev.\ Lett.\  } {\bf 97} (2006) 131801
  [hep-ex/0602020].

\bibitem{Pospelov:2005pr}
  M.~Pospelov and A.~Ritz,
  {\it Electric dipole moments as probes of new physics},
  {\it Annals Phys.\ } {\bf 318} (2005) 119
  [hep-ph/0504231].

\bibitem{LaGo}
  S.~K.~Lamoreaux and R.~Golub,
  {\it Experimental searches for the neutron electric dipole moment},
  {\it J.\ Phys.\ } {\bf G 36} (2009) 104002.

\bibitem{Farley:2003wt}
  F.~J.~M.~Farley, K.~Jungmann, J.~P.~Miller, W.~M.~Morse, Y.~F.~Orlov, B.~L.~Roberts, Y.~K.~Semertzidis and A.~Silenko {\it et al.},
  {\it A New method of measuring electric dipole moments in storage rings},
  {\it Phys.\ Rev.\ Lett.\ } {\bf 93} (2004) 052001
  [hep-ex/0307006].

\bibitem{Semertzidis:2011qv}
  Y.~K.~Semertzidis [Storage Ring EDM Collaboration],
  {\it A Storage Ring proton Electric Dipole Moment experiment: most sensitive experiment to CP-violation beyond the Standard Model},
  arXiv:1110.3378 [physics.acc-ph].

\bibitem{Rathmann:2011zz}
  F.~Rathmann and N.~Nikolaev,
  {\it Precursor experiments to search for permanent electric dipole moments (EDMs) of protons and deuterons at COSY},
  {\it PoS } {\bf STORI11} (2011) 029.

\bibitem{Lehrach:2012eg}
  A.~Lehrach, B.~Lorentz, W.~Morse, N.~Nikolaev and F.~Rathmann,
  {\it Precursor experiments to search for permanent electric dipole moments (EDMs) of protons and deuterons at COSY},
  arXiv:1201.5773 [hep-ex].

\bibitem{Aoki:1989rx}
  S.~Aoki and A.~Gocksch,
  {\it The neutron electric dipole moment in lattice QCD},
  {\it Phys.\ Rev.\ Lett.\ } {\bf 63} (1989) 1125
   [{\it Erratum ibid} {\bf 65} (1990) 1172].

\bibitem{Aoki:1990ix}
  S.~Aoki, A.~Gocksch, A.~V.~Manohar and S.~R.~Sharpe,
  {\it Calculating the neutron electric dipole moment on the lattice},
  {\it Phys.\ Rev.\ Lett.\ } {\bf 65} (1990) 1092.

\bibitem{Shintani:2006xr}
  E.~Shintani, S.~Aoki, N.~Ishizuka, K.~Kanaya, Y.~Kikukawa, Y.~Kuramashi, M.~Okawa and A.~Ukawa et al.,
  {\it Neutron electric dipole moment with external electric field method in lattice QCD},
  {\it Phys.\ Rev.\ } {\bf D 75} (2007) 034507
  [hep-lat/0611032].

\bibitem{Shintani:2008nt}
  E.~Shintani, S.~Aoki and Y.~Kuramashi,
  {\it Full QCD calculation of neutron electric dipole moment with the external electric field method},
  {\it Phys.\ Rev.\ } {\bf D 78} (2008) 014503
  [arXiv:0803.0797 [hep-lat]].

\bibitem{Shintani:2005xg}
  E.~Shintani, S.~Aoki, N.~Ishizuka, K.~Kanaya, Y.~Kikukawa, Y.~Kuramashi, M.~Okawa and Y.~Tanigchi et al.,
  {\it Neutron electric dipole moment from lattice QCD},
  {\it Phys.\ Rev.\ } {\bf D 72} (2005) 014504
  [hep-lat/0505022].

\bibitem{Berruto:2005hg}
  F.~Berruto, T.~Blum, K.~Orginos and A.~Soni,
  {\it Calculation of the neutron electric dipole moment with two dynamical flavors of domain wall fermions},
  {\it Phys.\ Rev.\ } {\bf D 73} (2006) 054509
  [hep-lat/0512004].


\bibitem{Izubuchi:2008mu}
  T.~Izubuchi, S.~Aoki, K.~Hashimoto, Y.~Nakamura, T.~Sekido and G.~Schierholz,
  {\it Dynamical QCD simulation with theta terms},
  {\it PoS} {\bf LAT 2007} (2007) 106
  [arXiv:0802.1470 [hep-lat]].

\bibitem{Aoki:2008gv}
  S.~Aoki, R.~Horsley, T.~Izubuchi, Y.~Nakamura, D.~Pleiter, P.~E.~L.~Rakow, G.~Schierholz and J.~Zanotti,
  {\it The Electric dipole moment of the nucleon from simulations at imaginary vacuum angle theta},
  arXiv:0808.1428 [hep-lat].

\bibitem{confX}
  E.~Shintani,
  talk given at {\it the Xth Quark Confinement and the Hadron Spectrum}, Garching, Oct. 8--12, 2012.


\bibitem{Gerrit}
  G.~Schierholz,
  talk given at {\it the ECT* Workshop on EDM Searches at Storage Rings}, Trento,
  Oct. 2--5, 2012

\bibitem{Brodsky:2006ez}
  S.~J.~Brodsky, S.~Gardner and D.~S.~Hwang,
  {\it Discrete symmetries on the light front and a general relation connecting nucleon electric dipole and anomalous magnetic moments},
  {\it Phys.\ Rev.\ } {\bf D 73} (2006) 036007
  [hep-ph/0601037].


\bibitem{Liu:2008gr}
  K.-F.~Liu,
  {\it Neutron electric dipole moment at fixed topology},
  {\it Mod.\ Phys.\ Lett.\ } {\bf A 24} (2009) 1971
  [arXiv:0807.1365 [hep-ph]].

\bibitem{Mereghetti:2010tp}
  E.~Mereghetti, W.~H.~Hockings and U.~van Kolck,
  {\it The effective chiral Lagrangian from the theta term},
  {\it Annals Phys.\ } {\bf 325} (2010) 2363
  [arXiv:1002.2391 [hep-ph]].

\bibitem{Thomas:1994wi}
  S.~D.~Thomas,
  {\it Electromagnetic contributions to the Schiff moment},
  {\it Phys.\ Rev.\ } {\bf D 51} (1995) 3955
  [arXiv:hep-ph/9402237].

\bibitem{Borasoy:2000pq}
  B.~Borasoy,
  {\it The electric dipole moment of the neutron in chiral perturbation  theory},
  {\it Phys.\ Rev.\ } {\bf D 61} (2000) 114017
  [arXiv:hep-ph/0004011].

\bibitem{Crewther:1979pi}
  R.~J.~Crewther, P.~Di Vecchia, G.~Veneziano and E.~Witten,
  {\it Chiral estimate of the electric dipole moment of the neutron in quantum
  chromodynamics},
  {\it Phys.\ Lett.\ } {\bf B 88} (1979) 123
  [{\it Erratum ibid} {\bf B 91} (1980) 487].

\bibitem{Pich:1991fq}
  A.~Pich and E.~de Rafael,
  {\it Strong CP violation in an effective chiral Lagrangian approach},
  {\it Nucl.\ Phys.\ } {\bf B 367} (1991) 313.

\bibitem{Narison:2008jp}
  S.~Narison,
  {\it A fresh look into the neutron EDM and magnetic susceptibility},
  {\it Phys.\ Lett.\ } {\bf B 666} (2008) 455
  [arXiv:0806.2618 [hep-ph]].

\bibitem{Hockings:2005cn}
  W.~H.~Hockings and U.~van Kolck,
  {\it The electric dipole form factor of the nucleon},
  {\it Phys.\ Lett.\ } {\bf B 605} (2005) 273
  [arXiv:nucl-th/0508012].

\bibitem{Ottnad:2009jw}
  K.~Ottnad, B.~Kubis, U.-G.~Mei{\ss}ner and F.-K.~Guo,
  {\it New insights into the neutron electric dipole moment},
  {\it Phys.\ Lett.\ } {\bf B 687} (2010) 42
  [arXiv:0911.3981 [hep-ph]].

\bibitem{Mereghetti:2010kp}
  E.~Mereghetti, J.~de Vries, W.~H.~Hockings, C.~M.~Maekawa and U.~van Kolck,
  {\it The electric dipole form factor of the nucleon in chiral perturbation theory to sub-leading order},
  {\it Phys.\ Lett.\ } {\bf B 696} (2011) 97
  [arXiv:1010.4078 [hep-ph]].

\bibitem{O'Connell:2005un}
  D.~O'Connell and M.~J.~Savage,
  {\it Extrapolation formulas for neutron EDM calculations in lattice QCD},
  {\it Phys.\ Lett.\ } {\bf B 633} (2006) 319
  [hep-lat/0508009].

\bibitem{Chen:2007ug}
  J.-W.~Chen, D.~O'Connell and A.~Walker-Loud,
  {\it Universality of mixed action extrapolation formulae},
  {\it JHEP } {\bf 04} (2009) 090
  [arXiv:0706.0035 [hep-lat]].

\bibitem{Donoghue}
  J.~F.~Donoghue, E.~Golowich and B.~R.~Holstein,
  {\it Dynamics of the Standard Model},
  Cambridge University Press, Cambridge (1992).

\bibitem{Gasser:1984gg}
  J.~Gasser and H.~Leutwyler,
  {\it Chiral perturbation theory: Expansions in the mass of the strange quark},
  {\it Nucl.\ Phys.\ } {\bf B 250} (1985) 465.

\bibitem{Leutwyler:1996sa}
  H.~Leutwyler,
  {\it Bounds on the light quark masses},
  {\it Phys.\ Lett.\ } {\bf B 374} (1996) 163
  [arXiv:hep-ph/9601234].

\bibitem{HerreraSiklody:1996pm}
  P.~Herrera-Siklody, J.~I.~Latorre, P.~Pascual and J.~Taron,
  {\it Chiral effective Lagrangian in the large-$N_c$ limit: The nonet case},
  {\it Nucl.\ Phys.\ } {\bf B 497} (1997) 345
  [arXiv:hep-ph/9610549].

\bibitem{Ottnadthesis}
  K.~Ottnad, {\it The electric dipole form factor of the neutron in chiral perturbation theory},
  Diploma thesis, University of Bonn (2009).

\bibitem{Becher:1999he}
  T.~Becher and H.~Leutwyler,
  {\it Baryon chiral perturbation theory in manifestly Lorentz invariant form},
  {\it Eur.\ Phys.\ J.\ } {\bf C 9} (1999) 643
  [hep-ph/9901384].


\bibitem{Borasoy:1996bx}
  B.~Borasoy and U.-G.~Mei{\ss}ner,
  {\it Chiral expansion of baryon masses and sigma terms},
  {\it Annals Phys.\ } {\bf 254} (1997) 192
  [hep-ph/9607432].

\bibitem{Bsaisou:2012rg}
  J.~Bsaisou, C.~Hanhart, S.~Liebig, U.-G.~Mei{\ss}ner, A.~Nogga and A.~Wirzba,
  {\it The electric dipole moment of the deuteron from the QCD $\theta$-term},
  arXiv:1209.6306 [hep-ph].

\bibitem{PDG}
  J. Beringer {\it et al.} [Particle Data Group],
  {\it Review of Particle Physics},
  {\it Phys.\ Rev.\ } {\bf D 86} (2012) 010001.

\bibitem{HerreraSiklody:1997kd}
  P.~Herrera-Siklody, J.~I.~Latorre, P.~Pascual and J.~Taron,
  {\it $\eta-\eta'$ mixing from $U(3)_L\bigotimes U(3)_R$ chiral perturbation theory},
  {\it Phys.\ Lett.\ } {\bf B 419} (1998) 326
  [hep-ph/9710268].

\bibitem{Tiburzi:2007ep}
  B.~C.~Tiburzi,
  {\it External momentum, volume effects, and the nucleon magnetic moment},
  {\it Phys.\ Rev.\ } {\bf D 77} (2008) 014510
  [arXiv:0710.3577 [hep-lat]].

\bibitem{Greil:2011aa}
  L.~Greil, T.~R.~Hemmert and A.~Schafer,
  {\it Finite Volume Corrections to the Electromagnetic Current of the Nucleon},
  {\it Eur.\ Phys.\ J.\ } {\bf A 48} (2012) 53
  [arXiv:1112.2539 [hep-ph]].

\bibitem{Gasser:1987zq}
  J.~Gasser and H.~Leutwyler,
  {\it Spontaneously broken symmetries: Effective lagrangians at finite volume},
  {\it Nucl.\ Phys.\ } {\bf B 307} (1988) 763.

\bibitem{Geng:2011wq}
  L.-s.~Geng, X.-l.~Ren, J.~Martin-Camalich and W.~Weise,
  {\it Finite-volume effects on octet-baryon masses in covariant baryon chiral perturbation theory},
  {\it Phys.\ Rev.\ } {\bf D 84}  (2011) 074024
  [arXiv:1108.2231 [hep-ph]].

\bibitem{Gasser:1986vb}
  J.~Gasser and H.~Leutwyler,
  {\it Light quarks at low temperatures},
  {\it Phys.\ Lett.\ } {\bf B 184} (1987) 83.

\bibitem{Tiburzi:2008pa}
  B.~C.~Tiburzi,
  {\it Volume effects for pion two-point functions in constant electric and magnetic fields},
  {\it Phys.\ Lett.\ } {\bf B 674} (2009) 336
  [arXiv:0809.1886 [hep-lat]].


\bibitem{Gasser:1987rb}
  J.~Gasser, M.~E.~Sainio and A.~\v{S}varc,
  {\it Nucleons with chiral loops},
  {\it Nucl.\ Phys.\ } {\bf B 307} (1988) 779.

\bibitem{Jenkins:1990jv}
  E.~E.~Jenkins and A.~V.~Manohar,
  {\it Baryon chiral perturbation theory using a heavy fermion Lagrangian},
  {\it Phys.\ Lett.\ } {\bf B 255} (1991) 558.

\bibitem{Bernard:1995dp}
  V.~Bernard, N.~Kaiser and U.-G.~Mei{\ss}ner,
  {\it Chiral dynamics in nucleons and nuclei},
  {\it Int.\ J.\ Mod.\ Phys.\ } {\bf E 4} (1995) 193
  [hep-ph/9501384].

\bibitem{Bernard:2007zu}
  V.~Bernard,
  {\it Chiral perturbation theory and baryon properties},
  {\it Prog.\ Part.\ Nucl.\ Phys.\ } {\bf 60} (2008) 82
  [arXiv:0706.0312 [hep-ph]].

\bibitem{Ellis:1997kc}
  P.~J.~Ellis and H.-B.~Tang,
  {\it Pion nucleon scattering in a new approach to chiral perturbation theory},
  {\it Phys.\ Rev.\ } {\bf C 57} (1998) 3356
  [hep-ph/9709354].

\bibitem{Beane:2004tw}
  S.~R.~Beane,
  {\it Nucleon masses and magnetic moments in a finite volume},
  {\it Phys.\ Rev.\ } {\bf D 70} (2004) 034507
  [hep-lat/0403015].

\bibitem{handbook}
  M.~Abramowitz and I.~A.~Stegun,
  {\it Handbook of Mathematical Functions with Formulas, Graphs, and Mathematical Tables},
  Dover, New York (1972).

\end{thebibliography}
\end{document}